\documentclass{aa}
\usepackage[varg]{txfonts}
\usepackage[colorlinks=true, citecolor=blue, linkcolor=blue]{hyperref}
\usepackage{subfig}
\usepackage{natbib}
\usepackage{multirow}
\bibpunct{(}{)}{;}{a}{}{,}

\usepackage{graphicx}
\begin{document}
\title{An optical spectroscopic and polarimetric study of the microquasar binary system SS 433}
\author{Paolo Picchi\inst{1} \and Steven N. Shore\inst{1,2} \and Eamonn J. Harvey\inst{3} \and Andrei Berdyugin\inst{4} }
\institute{
Dipartimento di Fisica "Enrico Fermi'', Universit\`a di Pisa; \email{steven.neil.shore@unipi.it}
\and 
INFN- Sezione Pisa, largo B. Pontecorvo 3, I-56127 Pisa, Italy
\and
Astrophysics Research Institute, Liverpool John Moores University, IC2, LSP, 146 Brownlow Hill, Liverpool L3 5RF, UK
\and
Department of Physics and Astronomy, FI-20014 University of Turku, Finland
 }

\date{Submitted: --  ; accepted ---}


\abstract
{}{Our aim is to study the mass transfer, accretion environment, and wind outflows in the SS 433 system, concentrating on the so-called stationary lines.}{We used archival high-resolution (XSHOOTER) and low-resolution (EMMI) optical spectra, new optical multi-filter polarimetry, and low-resolution optical spectra (Liverpool Telescope), spanning an interval of a decade and a broad range of precessional and orbital phases, to derive the dynamical properties of the system.}{Using optical interstellar absorption lines and H I 21 cm profiles, we derive E(B-V) =0.86$\pm$0.10, with an upper limit of E(B-V)=1.8$\pm$0.1 based on optical Diffuse Interstellar Bands.  We obtain  revised values for the ultraviolet and U band polarizations and polarization angles (PA), based on a new calibrator star at nearly the same distance as SS 433 that corrects the published measurement and yields the same position angle (PA) as the optical.  The polarization wavelength dependence is consistent with optical-dominating electron scattering with a Rayleigh component in U and the UV filters.  No significant phase modulation was found for PA while there is significant variability in the polarization level. We fortuitously caught a flare event; no polarization changes were observed but we confirm the previously reported associated emission line variations. Studying profile modulation of multiple lines of H I, He I, O I, Na I, Si II, Ca II, Fe II with precessional and orbital phase, we derive properties for the accretion disk and present evidence for a strong disk wind, extending published results.  Using transition-dependent systemic velocities, we probe the velocity gradient of the wind, and demonstrate that it is also variable on timescales unrelated to the orbit.  Using the rotational velocity, around 140$\pm$20 km s$^{-1}$, a redetermined mass ratio $q = 0.37\pm0.04$, and masses M$_X=4.2\pm0.4$M$_\odot$, M$_A = 11.3\pm0.6$M$_\odot$, the radius of the A star fills -- or slightly overfills -- its Roche surface.  We devote particular attention to the O I 7772 \AA\ and 8446 \AA\ lines, finding that they show different but related orbital and precessional modulation and there is no evidence for a circumbinary component.  The spectral line profile variability can, in general, be understood with an ionization stratified outflow predicted by thermal wind modeling, modulated by different lines of sight through the disk produced by its precession.  The wind can also account for an extended equatorial structure detected at long wavelength.}{}
\maketitle

\section{Introduction}
\label{sec:int}
As the prototype Galactic microquasar, the binary system SS 433 (=V1343 Aql) has been extensively studied from TeV through low frequency radio and spatial resolutions down to the mas scale since its discovery. Much of this work has concentrated on its iconic jets and their phenomenology, the standard kinematic model for which is the precession of a tilted accretion disk around a compact, relativistic object~\citep[see][for a review]{fabrika:review} with a period of about 162 days. The jets move ballistically at about 0.26c into the surrounding supernova remnant W50.  Their geometry and kinematics have been measured over decades since the first interferometric imaging in the 1980s~\citep{hjellming:jets}. The inferred accretion rates are super-Eddington~\citep[see][]{fuchs:accretion, shkovskii:accretionrate, kotani:ironline, fabrika:supercritical} so the disk is likely thick and losing mass, with the supply of material coming from the companion that is presumed to fill (or overfill) its Roche surface~\citep{hillwig:one}.  Efforts to unravel the nature of the gainer have concentrated on obtaining its mass, based mainly on emission line studies~\citep{fabrika:massfunction, fabrika:supercritical}.  The system is photometrically variable due to the partial eclipse of the gainer by the companion on a 13 day orbit.  The photospheric lines of the loser~\citep{hillwig:one} have been detected and it is identified as an early A supergiant. Combining the mass function, mass ratio, and eclipse solution, the range of masses for the gainer is consistent with a stellar mass black hole. The greatest attention has been paid, to date, on the inner portion of the accretion disk from which the jets emerge and that provide insights into processes under strong gravity that mimic active galactic nuclei. In this paper, in contrast, we focus on the accretion disk and its associated wind.  Our aim is to present a unified description of the hydrodynamic processes inferred from low and high resolution optical spectroscopy and filter polarimetry.

\section{Observations}
\label{sec:observations}
The dataset used in this work combines new and archival material. The full journal of observations  is presented in Appendix~\ref{app:journal}.  The new observations  are multi-filter polarimetry and low-resolution spectroscopy, taken at the Liverpool Telescope (LT, La Palma, Spain) and the T60 telescope (Haleakala, Hawaii, USA), both operated remotely. Polarimetry data were obtained with the RINGO3 (LT) and Dipol-2 polarimeters (T60). The instruments simultaneously measure linear polarization in the B, V, and R bands~\citep{arnold:ringo3, piirola:dipol}. Dipol-2 data calibration was performed by the telescope personnel~\citep[see][]{kosenkov:2017, piirola:cal}. The spectroscopic dataset comprises new low-resolution data~\citep[see][]{barnsley:frodospec}, acquired at the LT with the FRODOSpec spectrograph, and archival low- and high-resolution ESO spectra, obtained from NTT and VLT telescopes, with EMMI and XSHOOTER spectrographs  respectively\footnote{For details about the EMMI dataset see~\citet{blundell:burst} while the XSHOOTER data set corresponds to the ESO Program Id: 0101.D-0696 and P.I.: Waisberg, Idel~\citep[the dataset is published, in part, in][]{waisberg:collimated}.}, see Table~\ref{table:spectrographs}.
\begin{table}
\caption{Observation modes specifics of the spectrographs: the wavelength coverage, the resolving power R and the total number N of the spectra per each instrument.}
\label{table:spectrographs}
\centering
\begin{tabular}{cccc}
\hline\hline
Spectrograph & Wavelength range (nm) & R & N \\  
\hline  
FRODOSpec & 580-940 & 2200 & 11 \\
EMMI & 580-870 & 2727 & 62 \\
XSHOOTER & 300-559.5 & 4100 & 22 \\ 
XSHOOTER & 559.5-1024 & 8900 & 22 \\
\hline
\end{tabular}
\end{table}
Only the XSHOOTER spectra are flux-calibrated. The quantitative analysis presented here is based primarily on the EMMI and XSHOOTER spectra. The LT spectra were used for the study of an outburst detected during the observing campaign (see Sect.~\ref{sec:outburst}).
\section{Data analysis: ISM extinction}
\label{sec:analysis}
\subsection{ISM polarization contribution}
\label{subsec:pol}
The interstellar medium (ISM) component of the measured polarization was removed using comparison stars located in the same field of view of SS 433, as near as possible to the source in angle and  distance from the Gaia DR2 data. This permitted the derivation of the Serkowski curve in that region of the sky and the determination of the ISM Stokes parameters. The stars for which the polarization was measured are S2, S5, following the notation of \citet{leibowitz}, a new calibrator, SCal\footnote{2MASS id source: 19114434+0459264.}. S2 and S5 were measured with Dipol-2 while SCal, because of its magnitude, required Dipol-UF\footnote{For a description of the polarimeter, see \url{http://www.not.iac.es/instruments/dipol-uf/DIPol-UF_description.html}}, a new polarimeter installed at the Nordic Optical Telescope (NOT), that is better suited for faint objects. The polarization measurements of these stars are listed in Table~\ref{table:polStars}.
\begin{table}
\caption{Polarization and magnitude (Gaia bp filter) parameters of the field stars S2, S5 and SCal.}
\label{table:polStars}
\centering
\begin{tabular}{p{0.05\linewidth}p{0.08\linewidth}p{0.08\linewidth}p{0.15\linewidth}p{0.14\linewidth}p{0.12\linewidth}}
\hline\hline
\textbf{Star} & bp (mag) & Filter & P( $\%$ ) & PA (\degr) & HJD $-$ 2458319 \\  
\hline  
S2 & 13.66 & B & 1.91$\pm$0.06 & 49.2$\pm$1.0 & 0.55 \\
     & & V & 1.86$\pm$0.06 & 48.0$\pm$0.9 & 0.55 \\
     & & R & 1.78$\pm$0.03 & 48.0$\pm$0.4 & 0.55 \\
S5 & 15.67& B & 1.10$\pm$0.20 & 51.7$\pm$5.2 &  0.40 \\
     & & V & 1.58$\pm$0.21 & 52.6$\pm$3.7 &  0.40\\
     & & R & 1.27$\pm$0.09 & 49.7$\pm$2.0 & 0.40\\
SCal & 16.43 & B & 4.29$\pm$0.20 & 176.3$\pm$1.3 & 369.04\\
        & & V & 4.37$\pm$0.07 & 176.1$\pm$0.5 & 369.04\\
       & & R & 4.46$\pm$0.03 & 176.9$\pm$0.2 & 369.04\\
\hline
\end{tabular}
\end{table}
The data in the three filters were fit with the Serkowski law~\citep{serkowski:ism}. The measures for S2 and S5 are nearly the same in both the Polarization Level (PL) and  the Polarization Angle (PA) (see Table~\ref{table:polStars}), suggesting that the stars are intrinsically unpolarized and their polarization is probably only interstellar in origin. SCal, instead, indicates how the ISM induced polarization varies as the distance increases. This source is the physically nearest to SS 433.   
The fit parameters are listed in Table~\ref{table:Serkowski}, together with the parallaxes measured by Gaia, which indicate that the stars are respectively located at about $1.1$, $0.7$ and 4 kpc from the Earth. SS 433 has a parallax of $p_{SS433} = 0.22\pm0.06$, which corresponds to a distance of about $4.5$ kpc. 
Using only S2 and S5 and interpolating the Serkowski $p_{max}$ up to the SS 433 parallax, gives $P^{SS433}_{max} = 2.79\pm0.26$, which is significantly lower than the value given by SCal. Moreover, the PA changes drastically. Thus, removing the interstellar contribution requires measuring the polarization of calibrators located spatially as near as possible to the source. For these reasons, SCal was used to obtain the intrinsic polarization values of SS 433.
\begin{table}
\caption{Parallaxes and Serkowski curves fitted.}
\centering
\begin{tabular}{ccccc}
\hline\hline
\textbf{Star} & Parallax (mas) & P$_{max}$ & $\lambda_{max}$ (\AA) & $\frac{\chi^2}{\nu}$  \\
\hline
S2 & 0.98$\pm$0.03 & 1.93$\pm$0.04 & 5082.5$\pm$192.2 & 0.82 \\      
\\
S5 & 1.51$\pm$0.06 & 1.33$\pm$0.08 & 5786$\pm$1059  & 1.98 \\      
\\
SCal & 0.25$\pm$0.08 & 4.48$\pm$0.03 & 6065.3$\pm$207.0 & 3.49 \\
\hline
\label{table:Serkowski}\\
\end{tabular}
\end{table}
The interstellar Stokes parameters of the respective filters are provided in Table~\ref{table:ismstokes}. The final Stokes parameters in the B, V, R filters were obtained using SCal while those in the U filter were obtained from the Serkowski law fit assuming the same PA of the B filter.
\begin{table}
\caption{SCal ISM Stokes parameters in the various filters. The Stokes parameters in the B, V, R bands were directly measured by Dipol-UF while the values in the U band were derived from the relative Serkowski curve.}  
\centering
\begin{tabular}{cccccc}
\hline\hline
Filter & q ($\%$) & u($\%$) \\
\hline U & 3.21$\pm$0.20 & $-0.42\pm$0.23 \\
 B & 4.26$\pm$0.20 & $-0.55\pm$0.19 \\
 V & 4.33$\pm$0.07 & $-0.60\pm$0.08 \\
 R & 4.43$\pm$0.03 & $-0.48\pm$0.03 \\		
\hline
\label{table:ismstokes}\\
\end{tabular}
\end{table}
\subsection{ISM $A_V$ and $E(B-V)$ determination}
\label{sec:extinction}
The total extinction coefficient, $A_V$, was first derived by~\citet{wagner:reddening} by fitting blackbody curves to  optical spectrophotometric observations. He obtained $A_V=7.8\pm0.5$ mag.  We derived a range for $E(B-V)$ and a value for $A_V$, from two independent methods: (1) H I column density along the SS 433 line of sight and (2) the equivalent width (EW) of DIBs present in the spectra. Both are based on measurements of tracers that are external to SS 433 and do not require any assumptions regarding the system's spectrum. 

\citet{guver:ism} obtained a calibration between the H I column density and the extinction,
\begin{displaymath}
\label{eqn:extinctionH}
\text{N}_\text{H} ( \text{cm}^{-2} ) = (2.21 \pm 0.09) \times 10^{21} A_V .
\end{displaymath}
The neutral hydrogen column density toward SS 433 is known from Galactic 21 cm profiles in the Leiden/Dwingeloo Survey for the sky north of $-30\degr$~\citep{kalberla:LAB}. We integrated the H I $21$ cm profile over the interstellar Na I D1 velocity range and divided it by the integral of the entire H I $21$ line to obtain the fraction of the total column density of H I toward that particular line of sight $N_{H,tot} = (6.40\pm0.28)\cdot 10^{21}$ cm$^{-2}$. For SS 433, we obtained N$_\text{H,SS}$=(5.95 $\pm$ 0.29)$\cdot 10^{21}$ cm$^{-2}$, A$_V = 2.69\pm0.17$ mag and $E(B-V) = 0.86\pm0.10$ mag using a total to selective extinction ratio of 3.1. This estimate is much lower than~\citet{wagner:reddening}, but agrees with the value obtained by~\citet{margon:two}, $A_V\sim2.3$. 

The DIBs are interstellar absorption transitions of still uncertain origin. Because of their ubiquity as a feature of the ISM, it is possible to derive $E(B-V)$ from their EW, which is correlated to the reddening~\citep{friedman:dibs, lan:dibs}. XSHOOTER spectra are used since they have the best signal to noise and resolution. We used the $5780$ \AA\ and $6614$ \AA\ bands. Two correlation laws have been used in this work: one, from~\citet{lan:dibs}, relates the EW to $E(B-V)$ through:
\begin{displaymath}
\label{eqn:dibs1}
EW = A\times(E(B-V))^\gamma,
\end{displaymath}
where $A$ and $\gamma$ are tabulated parameters that depend on the particular DIB considered. The other relation, from~\citet{friedman:dibs}, is linear (see Table $4$ of that paper for the correlation coefficients) and also depends on each DIB. We calculated the EWs for each of the 22 XSHOOTER spectra. The final values, shown in Table~\ref{table:dibs}, were obtained taking the mean, weighted with the inverse of the variances. Given a DIB, the algorithm selects two continuum regions, one blueshifted and the other redshifted respect to the line, closely adjacent to it. It then calculates the EW, choosing a point in the blueward and another in the redward intervals. Repeating the procedure for all the combinations of the points in the given intervals yields a distribution of EWs, from which mean and standard deviation are obtained. The width of the continuum intervals (varied within a few \AA s for each night) did not affect the mean EW, which remained consistent within the uncertainties. Table $5$ also lists the H I column density using~\citet{friedman:dibs} since they  found a better correlation with the H I column density (Fig. $2$ of their paper) than $E(B-V)$ (Fig. $4$). Although near the limit of the reliable range of the correlations, the measured EWs give an indication of the upper limit to both the column density and extinction. 
\begin{table*}
\caption{Reddening coefficients obtained from DIBs 5780 \AA\ and 6614 \AA. The footnotes indicate the calibration law used.}
\centering
\begin{tabular}{lcccc} 
\hline\hline
DIB & EW (m\AA) & E(B-V)\tablefootmark{a} & E(B-V)\tablefootmark{b} & H I (10$^{21}$ cm$^{-2}$)\tablefootmark{b} \\
\hline
5780 \AA& 641.3$\pm$0.9 &1.49$\pm$0.11& 1.26$\pm$0.01&4.35$\pm$1.37 \\
6614 \AA & 380.1$\pm$0.6&1.76$\pm$0.22& 1.78$\pm$0.02 &4.15$\pm$1.35\\
\hline
\label{table:dibs}\\
\end{tabular}
\tablefoot{
\tablefoottext{a}{\citet{lan:dibs}.}
\tablefoottext{b}{\citet{friedman:dibs}.}
}
\end{table*}

~\citet{gies:one} presented upper limits to the FUV (1150-1700 \AA) flux of SS 433 using the Hubble Space Telescope (HST) applying the extinction from~\citet{wagner:reddening} and~\cite{dolan:UV} (who measured the SS 433 flux at 2770 \AA\ with the High Speed Photometer (HSP) aboard HST) to extrapolate to the FUV. These UV observations further demonstrate that SS 433 is heavily reddened and that the $A_V$ and $E(B-V)$ values we obtained are consistent and model-independent.

\section{Data analysis: Spectroscopic variations}
\label{subsec:specanalysis}
The spectroscopic data used in this work were sufficient to distinguish intrinsic and modulated changes in precessional and orbital phases (for details about the data, see Appendix~\ref{app:journal}). The EMMI spectra covered a single, continuous sequence of observations. The system shows complex line profiles, which vary across transitions and in time (even from night to night), and modulated by the precession and orbital phases. Our analysis concentrates on the comparison of an ensemble of transitions from different species and arising from different processes to disentangle the line forming regions and dynamical environments. 

Spectroscopically, although SS 433 has been studied mainly for the jet lines,  the so called ``stationary lines'' that come from the binary system and trace the inner regions of the object and the accretion processes, leave many open questions. Usually, only a restricted set of lines have been studied, such as H$\alpha$, He II $4685$ \AA\ or the absorption features from the A-type supergiant~\citep{crampton:one,fabrika:massfunction,blundell:circ,hillwig:one,barnes:afs}, but SS 433 shows a wide variety of other transitions in the optical  that can  shed light on the binary environment. Some of these lines have already been reported, such as O I $8446$, $7774$ \AA, the He I triplets ($7065$, $5875$ \AA) and singlets ($6678$ \AA), and the Paschen lines, but their origin has been considered uncertain. It is not clear if they trace a region near the disk or if they come from more diffuse, circumstellar material. Most profiles for all species show both emission and absorption components, depending on the precessional phase, see Sect.~\ref{subsec:precmod}. For this reason, the spectra have been studied as a function of both precessional and orbital phases. The ephemerides used in this work are from~\citet{gora:ephem}:
\begin{displaymath}
\label{ephemeris1}
JD_{prec}= 2449998.0 + (162^{d}.278)\cdot E
\end{displaymath}
\begin{displaymath}
\label{eqn:ephemeris2}
JD_{orb}=  2450023.746+ (13^{d}.08223)\cdot E .
\end{displaymath}
These are photometric and set the time of maximum brightness during the precession and the time of center of eclipse, respectively. The orbital phase $\phi=0$ is defined by the eclipse of the disk by the supergiant.
\subsection{Precessional modulation}
\label{subsec:precmod}
\begin{figure*}
\centering
\includegraphics[width=17cm]{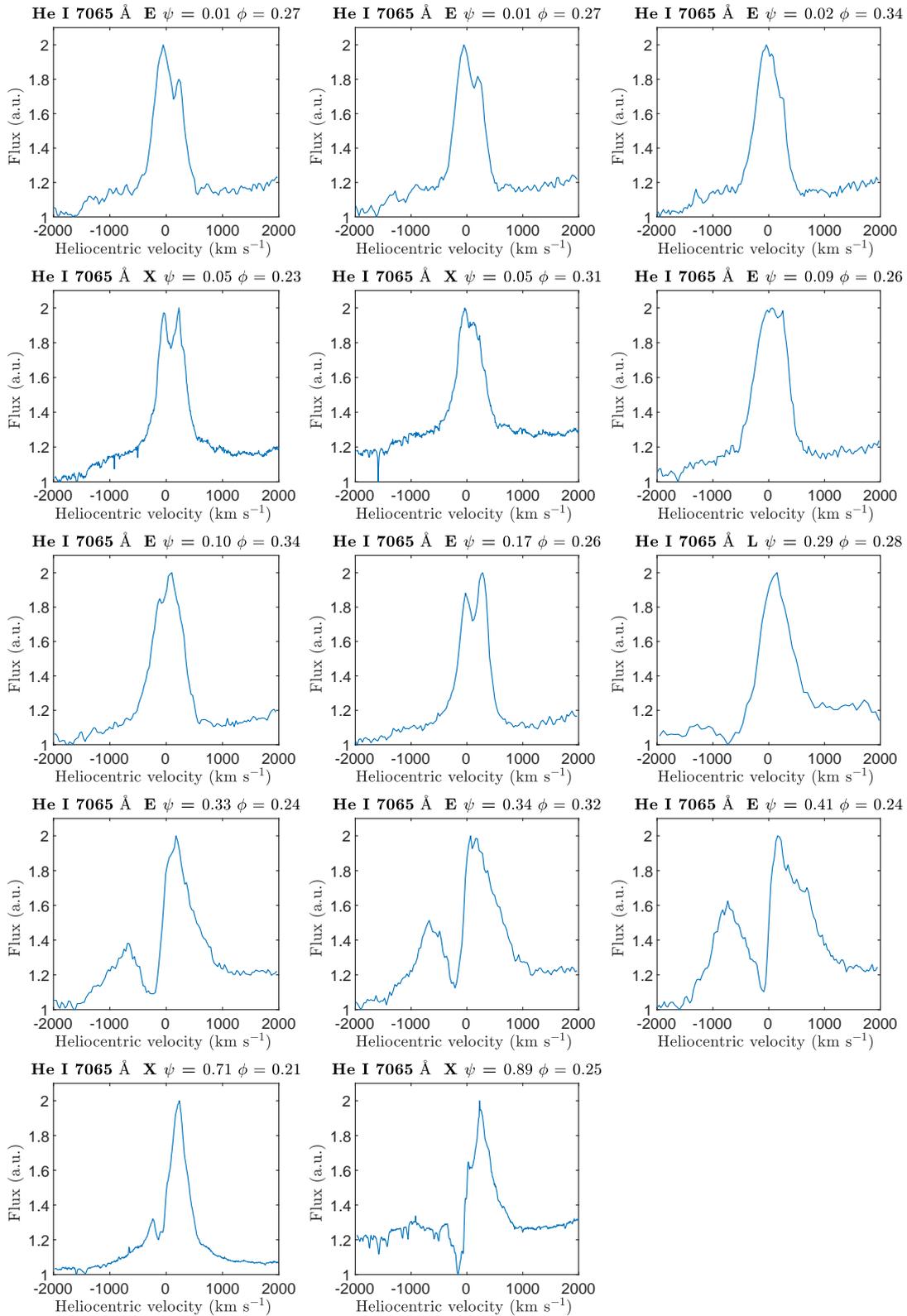}
\caption{He I $7065$ \AA\ line precessional sequence near the first elongation. The letters in each plot title indicate the spectrograph: ``X'' for XSHOOTER, ``E'' for EMMI (best phase coverage) and ``L'' for Liverpool.}
\label{fig:heprecmod1}
\end{figure*}

The precession alters the disk orientation to the line of sight. At phase $\psi = 0$  the disk is maximally face-on at $57\degr$ to the line of sight~\citep{fabrika:review}\footnote{Throughout the paper, the precessional phase is indicated with $\psi$ and the orbital with $\phi$.}. It is edge-on at phases $\psi\sim0.3$ and $\psi\sim0.6$. The lines that show absorption when the disk is edge-on all show the same modulation, so we concentrate on only one of them,  He I $7065$ \AA, since it is the strongest. The Balmer sequence is not displayed since it is present only in the XSHOOTER spectra, for which the precessional and orbital phase coverage are not sufficient, see appendix~\ref{app:journal}. H$\alpha$ is not shown because the EMMI data set is affected by an outburst~\citep{blundell:burst}. Fig.~\ref{fig:heprecmod1} shows the He I $7065$ \AA\ profile variations with the precessional phase at fixed orbital phase (near the first elongation, between $0.20$ and $0.35$). The letters in each plot title indicate the spectrograph: ``X'' for XSHOOTER, ``E'' for EMMI (best phase coverage) and ``L'' for Liverpool. The choice of the orbital phase does not affect the variations introduced by the disk precession. 
The absorption appears around $\psi=0.32$ and persists to  $\psi=0.48$ (last spectrum), so the medium that produces the absorption is in the plane of the disk and is quite extended. Moreover, it is asymmetric relative to the disk plane, since no absorption is detected at $\psi=0.2$ but it also persists  during the second half of the disk precession, as it is shown by the XSHOOTER spectra (see Sect.~\ref{subsec:afs}). The red wing extends to $+500$ km s$^{-1}$. The emission profile is asymmetric toward the red, especially when the disk is face-on and during primary eclipse; this is displayed by all the lines studied and is displayed in Fig.~\ref{fig:heorbmod0}.The absorption appears when the disk is edge-on and is always blue-shifted (Fig.~\ref{fig:heprecmod1}). This change in the profiles is long-term and systematic since it is present  in both the EMMI and  XSHOOTER spectra, although these were obtained more than ten years apart. 
\subsection{Orbital modulation}
\label{subsec:orbmod}
\begin{sidewaysfigure*}
\centering
\includegraphics[width=1\textwidth]{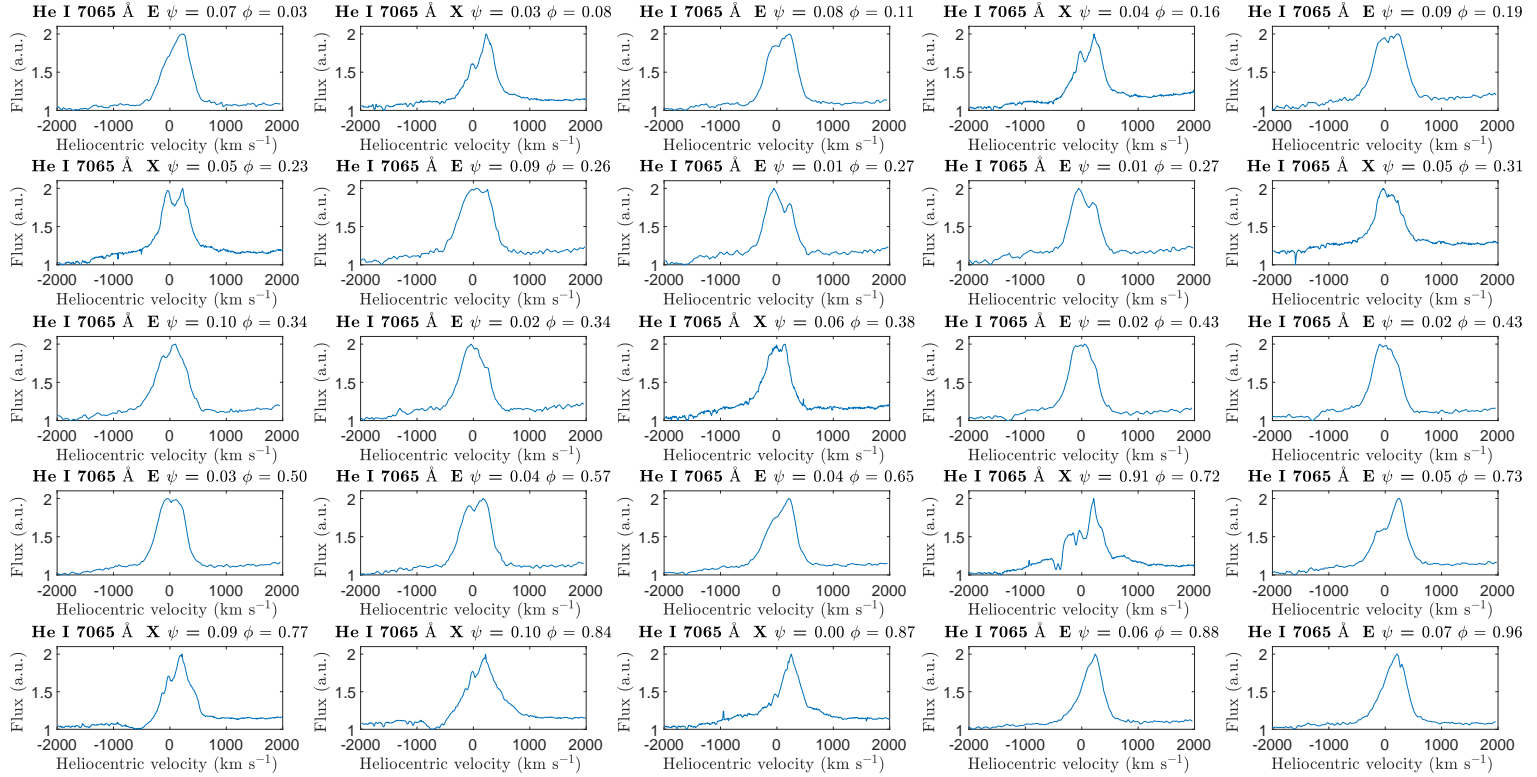}
\caption{He I $7065$ \AA\ line orbital sequence when the disk is face-on. The letters in each plot title indicate the spectrograph: ``X'' for XSHOOTER, ``E'' for EMMI (best phase coverage) and ``L'' for Liverpool.}
\label{fig:heorbmod0}
\end{sidewaysfigure*}

The orbital motion of the system produces variations in the line profiles that are less pronounced than those caused by precession. The orbital modulation is shown at nearly face-on phases ($0.90<\psi<0.10$) since the induced variations in the profiles are seen better and not affected by the disk absorption. The orbital modulation is more visible in the core of the profiles and in their two peak structure (since many lines show the double peak structure, the term ``blueshifted peak'' is replaced by \textsf{V} and ``redshifted peak'' by \textsf{R}). The He I $7065$ \AA\ line will again be the focus since it displays well the general characteristics shown by all the lines (i.~e., \textsf{V}/\textsf{R} on the core and the blueshifted absorption when the disk is edge-on). When the disk is face-on, in Fig.~\ref{fig:heorbmod0}, He I $7065$ \AA\ line shows only the \textsf{V}/\textsf{R} structure in the core. \textsf{R} is more shifted and prominent at primary eclipse, the double peak structure then emerges, with \textsf{V} maximum at $\phi\sim0.2-0.3$; \textsf{V}/\textsf{R} is symmetric at secondary eclipse. Finally, \textsf{V} decreases while \textsf{R} brightens. H$\alpha$, O I $8446$ \AA\ and all the other lines studied also show this pattern.

There is an asymmetry in velocity between the maximal redshift and blueshift. At $\phi\sim0$, \textsf{R} reaches $\sim+250$ km s$^{-1}$, while at phase $\sim0.3$ \textsf{V} is at around $-100$ to $-125$ km s$^{-1}$. The core of the lines traces a different region than the blueshifted absorption that appears only when the disk is edge-on (Fig.~\ref{fig:heprecmod1}). In particular, one spectrum in Fig.~\ref{fig:heorbmod0}, at precessional phase $\psi=0.91$, shows enhanced absorption near its core and detached blueshifted absorption at around $-400$ km s$^{-1}$. These multiple features are present in many other lines (such as Fe II 5169 \AA\ and Paschen series) but not in He II 4685 \AA. As we discuss in Sect.~\ref{subsec:afs}, these multiple absorptions give important information on the dynamical environment present in this system; they are visible at other intermediate phases, but only during the second half of the precession. The variations of the \textsf{V}/\textsf{R} structure of these profiles with orbital phase displays a strict correspondence with the He II line, which shows \textsf{V} more intense at $\phi\sim0.2$ and \textsf{R} more intense at $\phi\sim0$.
\subsection{Intrinsic line variations.}
\begin{figure*}
\centering
\includegraphics[width=17cm]{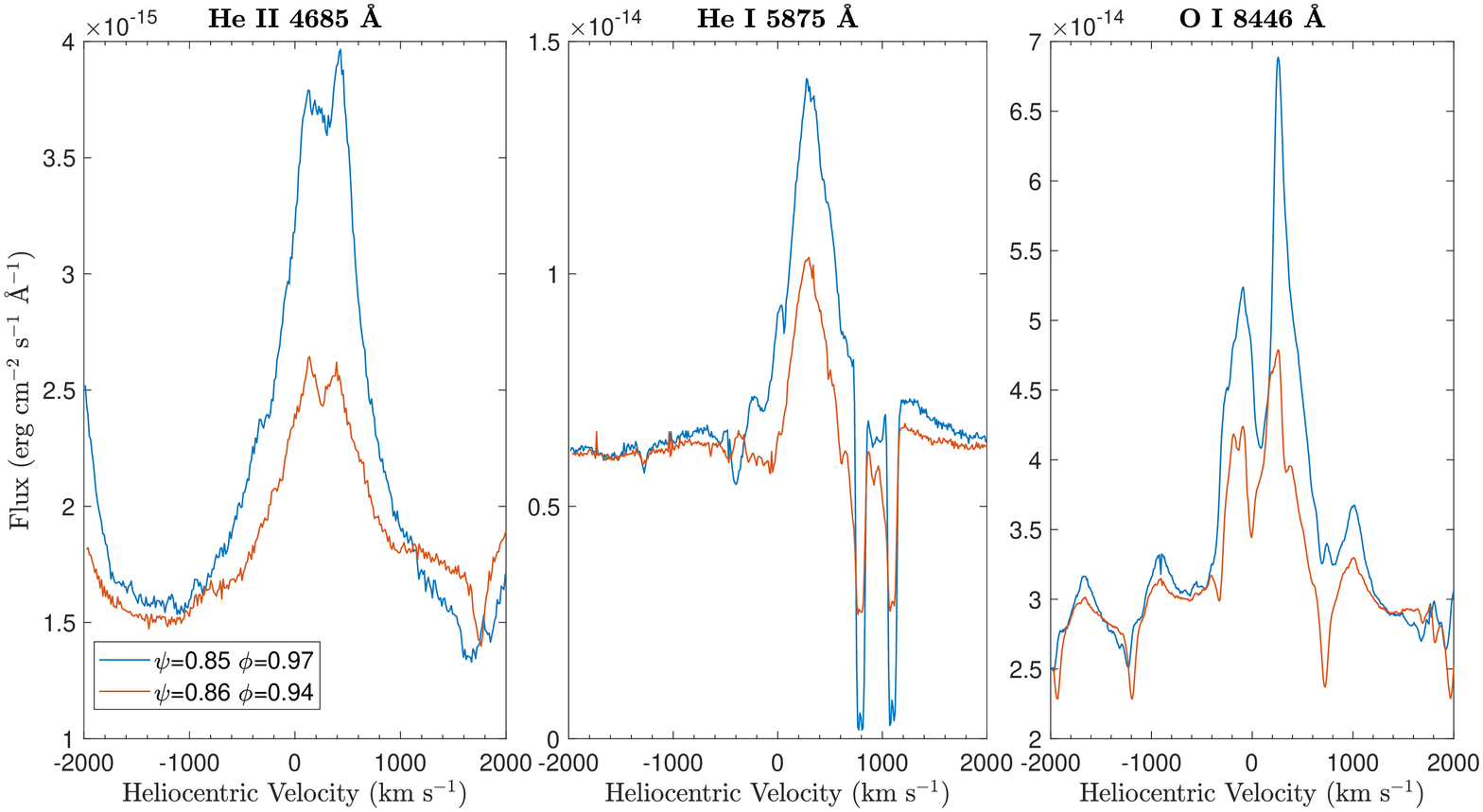}
\caption{Three different transitions at the same precessional and orbital phases. We note the clear differences in the profiles, and the blueshifted absorption features at negative velocities for He I 5875\AA. We note also the blueshifted absorption component of the Na I doublet in the He I $5875$ \AA\ subfigure. The lines have been plotted to the same continuum level for display.}
\label{fig:intrinsic_line_variations}
\end{figure*}

The high resolution, high S/N XSHOOTER spectra permitted an investigation of the intrinsic line profile variations of the system, that is, those that are due to neither  precession nor the orbit. We studied line profiles from different nights at precessional and orbital phases as close as possible. Fig.~\ref{fig:intrinsic_line_variations} shows the comparison between three transition: He II 4685 \AA, He I 5875 \AA\ and O I 8446 \AA.
\begin{figure}
\centering
\resizebox{\hsize}{!}{\includegraphics{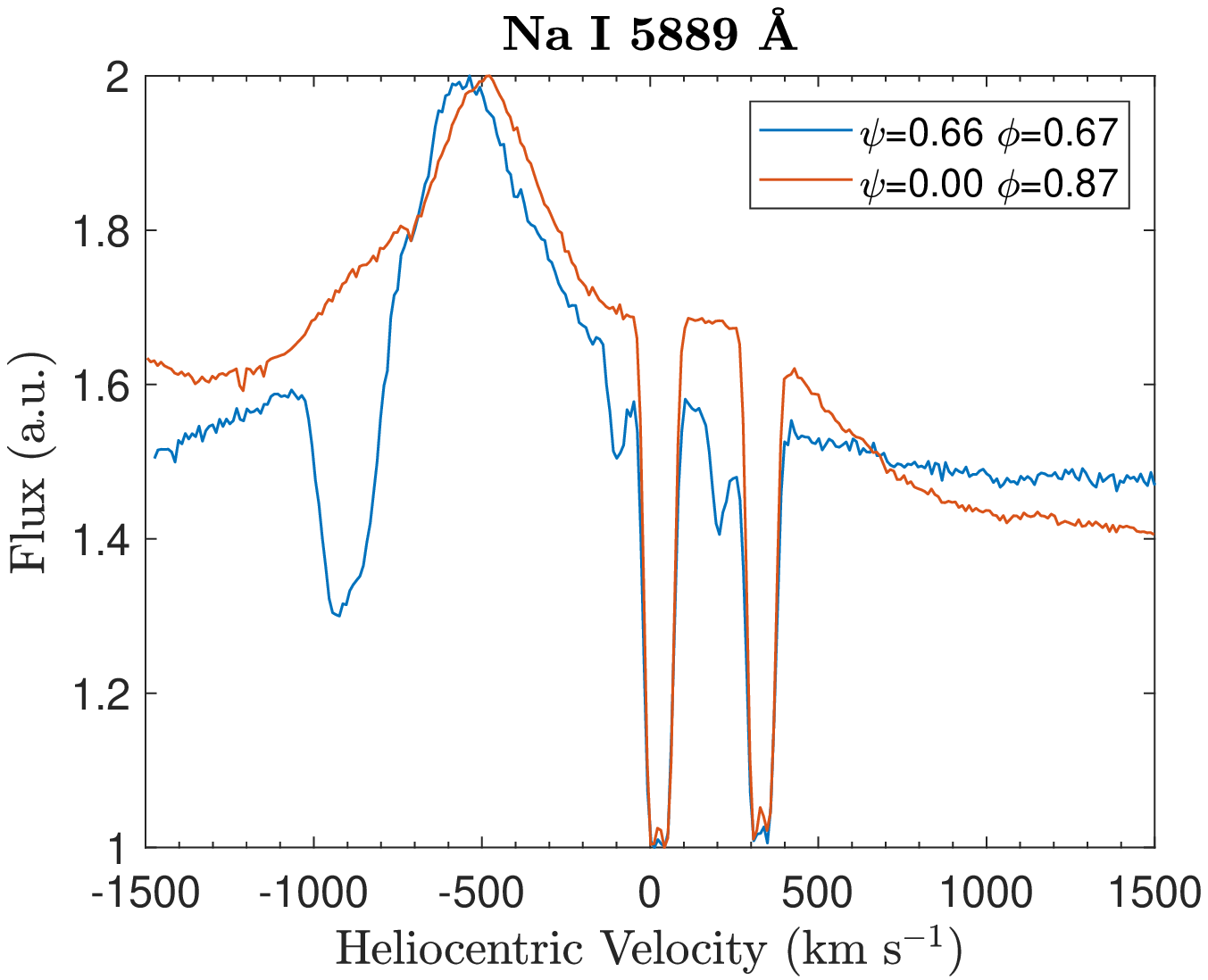}}
\caption{Two XSHOOTER spectra showing Na I D1  at opposite precessional phases. The appearance of additional absorption components at $\psi=0.66$ is important. The absorption was never detected at nearly face-on phases; it appears only when the disk is edge-on (also during the May sequence, $\psi\sim0.86$), indicating again that the disk is self-absorbing.}
\label{fig:nai}
\end{figure}
Although both phases are nearly identical (these are the only two spectra in the dataset that agree so closely in phase), there are clear differences in the line profiles. The He II 4686\AA\ profiles differ in the wings and peak ratios,  He I 5875 \AA\ at $\phi=0.97$ shows multiple, detached, weak absorption components while that at $\phi=0.94$ shows a single absorption at lower velocity.  In addition, there is a peculiar absorption of the Na I doublet in both the spectra (see Fig.~\ref{fig:nai}), detected for the first time in SS 433. The O I 8446 \AA\ displays changes in the peak relative intensity and multiple weak features, while the FWZI of the profile remains the same. Fig.~\ref{fig:intrinsic_line_variations} shows the transient changes in the flows present in the binary. All of these features can be detected \textit{only} at high resolution, highlighting it as a crucial requirement for any future spectroscopic observations of SS 433.  Without it, many features of the profiles will be undetectable and the analysis  affected by systematics.
\section{Results}
\label{sec:results}
This section presents the picture of SS 433 that emerges from the data analysis, concentrating on the properties of the thick accretion disk and flows in the system.
\subsection{Broadband filter polarimetry}
\label{sec:polresults}
The complete polarization data set is provided in Table~\ref{table:buncorr} to~\ref{table:runcorr}, which show the measured data, uncorrected for the ISM. 

\subsubsection{Photometric and polarimetric orbital variations.}
\label{subsec:pollight}
\begin{figure}
\centering
\resizebox{\hsize}{!}{\includegraphics{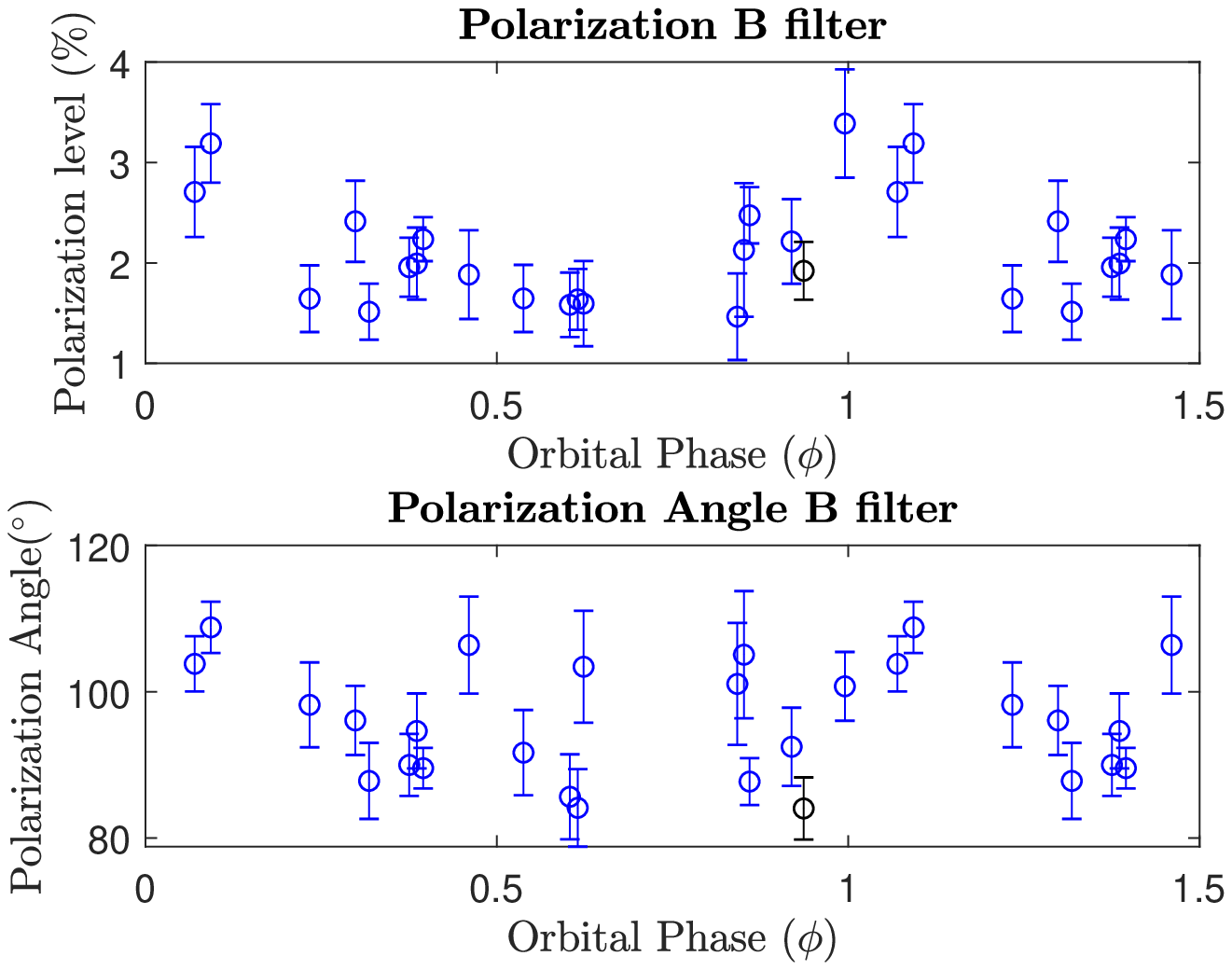}}
\caption[3pt]{Polarization light curve in the B filter. The outburst night is in black.}
\label{fig:sblightcurve}
\end{figure}

\begin{figure}
\centering
\resizebox{\hsize}{!}{\includegraphics{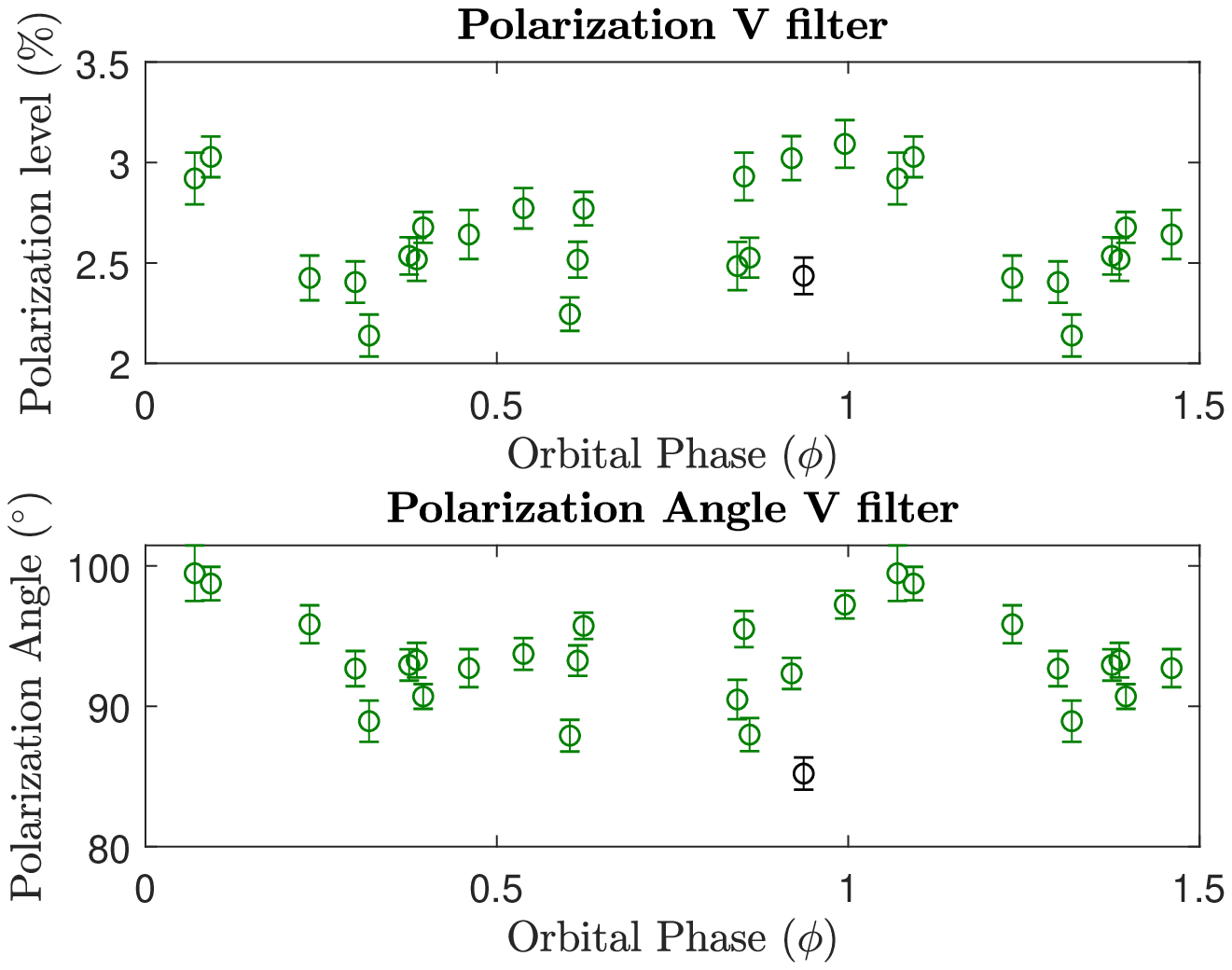}}
\caption[3pt]{Polarization light curve in the V filter. The outburst night is in black.}
\label{fig:svlightcurve}
\end{figure}

\begin{figure}
\centering
\resizebox{\hsize}{!}{\includegraphics{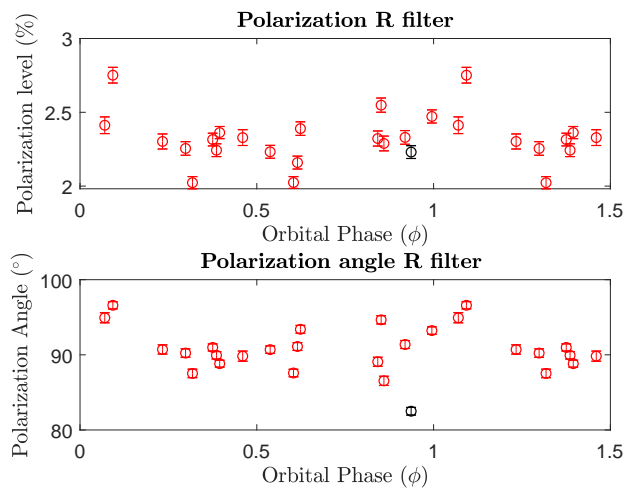}}
\caption[3pt]{Polarization light curve in the R filter. The outburst night is in black.}
\label{fig:srlightcurve}
\end{figure}

The orbit-modulated ISM corrected polarimetric curves for SS 433 are shown in Figs.~\ref{fig:sblightcurve},~\ref{fig:svlightcurve} and~\ref{fig:srlightcurve}.  The outburst night is plotted in black to distinguish it from the others. This event is not included in the discussion of the ``normal'' behavior of the system, but it will be considered in Sect.~\ref{sec:outburst}.
All filters show a significant change of PL and PA only around primary eclipse, when the disk is occulted by the supergiant. The B filter shows a stronger modulation in PL and PA over the orbit with respect to the other filters. The most important change is observed during primary eclipse. All bands have higher PL at $\phi=0$, while the PA decrease is phase shifted by $0.1$. To verify that the change of PL and PA over primary eclipse is significant, we compared the mean (weighted with the points dispersion) and standard deviation of the points near primary eclipse (specifically, between $0.9 \le \phi \le 0.1$) with the points located in the rest of the orbit, in all three filters. The Significance Level, SL $= (PL_E-PL_{OE})/\sigma_{OE}$, has also been calculated for both PL and PA. The subscript ``E'' refers to the points near primary eclipse while ``OE'' signifies those outside that interval (i.e., outside eclipse). The results are listed in Table~\ref{table:polstatistics}. The PL change during primary eclipse is more significant in the B and V filters and is higher in  B  (2.8 $\sigma$). The PA presents the same significance level in all three filters (1.5-1.8 $\sigma$). The PL variation around primary eclipse and the PA phase shift has not been reported previously since published studies have focused on the precession-related variation of the polarization, marginalized over the orbital modulation~\citep{efimov:ss433, mclean:ss433}. 
\begin{table}
\caption{Mean and standard deviations of PL and PA around primary eclipse and in the rest of the orbit. The Significance Level (SL), is shown too.}
\centering
\begin{tabular}{lcccc} 
\hline\hline
Filter \tablefootmark{a} & PL ($\%$) & SL$_\text{PL}$ & PA ($\deg$) & SL$_\text{PA}$ \\
\hline
B$_E$      & 2.84$\pm$0.53 & \multirow{2}{*}{2.8 $\sigma$} &103.2$\pm$6.9 & \multirow{2}{*}{1.5 $\sigma$} \\
B$_{OE}$   & 1.90$\pm$0.34 &                                               & 91.7$\pm$7.5  &                                                 \\
V$_E$         & 3.02$\pm$0.07 &\multirow{2}*{2.3 $\sigma$} & 96.4$\pm$3.2 & \multirow{2}*{1.5 $\sigma$} \\
V$_{OE}$   & 2.54$\pm$0.21 &                                               & 92.3$\pm$2.7 &                                                \\
 R$_{E}$    & 2.48$\pm$0.18 & \multirow{2}*{1.6 $\sigma$} & 94.1$\pm$2.2 & \multirow{2}*{1.8 $\sigma$} \\
 R$_{OE}$ & 2.26$\pm$0.14 &                                                & 90.1$\pm$2.2 &                                                 \\
\hline
\label{table:polstatistics}\\
\end{tabular}
\tablefoot{
\tablefoottext{a}{Eclipse = E. Outside Eclipse = OE.}
}
\end{table}
The phase dependent polarization variations are due to the compact object and its disk. One possibility is that although the disk is obscured, the scattering region is more extended and not as severely occulted. A contribution from the A star, by Rayleigh scattering around primary eclipse, could also be present. More detailed phase coverage with spectropolarimetric observations are needed to resolve this.  The R filter displays the lowest orbital modulation: the filter probes cooler regions in the periphery of the accretion disk, and, perhaps, more extended structures. 
Our polarization sequence (Table~\ref{table:buncorr} -~\ref{table:runcorr}), was obtained with the disk nearly edge-on.
Were we observing a classical thin disk, the polarization would depend strongly on the angle between the line of sight and the disk orientation, and the PL and PA variations should be very pronounced as the disk tilts. This does not appear to be the case, and suggests that either the disk is not geometrically thin, or that the polarization is produced by an extended structure. Since the spectra show self absorption of the disk during the edge-on phase, the polarization should come from a more extended, optically thin region, which is dynamically tied to the gainer and the disk (because of the orbital modulation of PL and PA) but its structure is different. A higher PL at $\phi\sim0$, in fact, can be due to dilution effects because of the eclipse of the disk or to the highlight of an extended coronal structure above the disk. The change in PL around primary eclipse is consistent with a continuum contribution from the disk coming with the  scattered (polarizing) structure.  Partial obscuration of the disk increases PL at primary eclipse.  The relative contribution is consistent with what appears to be required for interpreting the secondary absorption line spectrum (see below and Sect.~\ref{subsec:astar}).

\subsubsection{Redetermination of the ultraviolet polarization and the presence of a Rayleigh scatterer}
\label{sec:rayleigh}
Because of its  distance from the Earth and its location near the Galactic plane,  polarimetric measurements for SS 433 were strongly affected by the ISM systematics.
\citet{efimov:ss433} assumed a wavelength independent intrinsic polarization in their five filters (UBVRI), and obtained a fit for the ISM Serkowski curve. They took the time dependence of their data to be the intrinsic variation and separated it from the constant ISM values. To derive the ISM PA, they considered the line connecting the R and I filter points in the Stokes parameters, assuming that the slope they derived coincided with the ISM PA. They obtained $\theta\sim3^{\circ}$, which is significantly different than the value we obtained.

\citet{dolan:UV} measured the polarization level of SS 433 in the UBVR filters from the ground, and in the ultraviolet at $2770$ \AA\ (UV) using the polarimetric filters on the High Speed Photometer (HSP) aboard the Hubble Space Telescope (HST). They chose S$1$ and S$16$ as references, for which they measured a PA $\approx 3.5\degr$ and PL $\approx 2\%$.  After  correction, they found a PA nearly parallel to the jet direction in the UV and V filter, while in the B, R and I filters it is $\approx 140\degr$. Significantly, this difference between the B and V filters vanishes applying our correction (see Figs.~\ref{fig:sblightcurve}--\ref{fig:srlightcurve}.). They  also reported a marked increase in  PL from the U to the UV, $p_U = 8.9\pm0.9\%$ and $p_{UV}=18\pm3\%$, and proposed that this can be due to Rayleigh scattering.  Since Rayleigh and Thomson scattering have the same phase function, one would expect the same PA from the UV to the optical but, instead, they found PA $\approx 79\degr$  in the UV, $\approx 90\degr$ in U, and $\approx 140\degr$ in the visible. We obtain new intrinsic values for the U passband, applying our ISM correction: PL$_\mathrm{U}=(10.28\pm0.73) \%$ and PA$_\mathrm{U}=(90.9\pm2.2)^\circ$. The new U and corrected V  PAs show that the anomalies in~\citet{dolan:UV} are due only to their decontamination procedure. The V filter shows the same PA as the others, and, most important, the new PA derived in the U filter is consistent with the PAs measured in the present work. It appears that the polarization has two contributions: Thomson scattering, which dominates in the optical; and Rayleigh scattering, which comes from the same scattering region as the optical bandpasses. The jets have a position angle in the sky of $\sim100^\circ$, parallel to the PA in the U filter. Now, since the PL is always proportional to the optical depth $\tau$ in the optically thin limit, that is, $PL_\lambda \sim \tau_\lambda$. If the polarization arises from an extended disk, the PA must be orthogonal to it if the scatterer is optically thin. The measured polarization is consistent with the disk and not  the jets.  As noted by~\citet{dolan:UV}, the ratio of the UV to optical polarization should be $\approx 16$  from Rayleigh scattering alone.  The lower observed ratio implies that both Rayleigh and Thomson scattering are contributing in this source.  A further indication is that the ratio PL$_\text{E}$/PL$_\text{OE}$ progressively decreases from the B to the R filter (see Figs.~\ref{fig:sblightcurve}-\ref{fig:srlightcurve} and Table~\ref{table:polstatistics}). This change can be due to a dilution effect of the disk, which emission is more weighted toward the UV part of the spectrum. Moreover, a less effective contribution of the eclipse in the other bands is also possible, that is, the emission in V and R comes from a more extended region. 
\subsection{Spectroscopic results}
\label{sec:spectra}
\subsubsection{He II velocity curve}
\label{subsec:heIIorb}
Since its discovery, many attempts had been made to derive  masses for the  components of SS 433, but there remains a considerable dispersion among the results. The derived mass of the compact star ranges from 1.4 M$_{\odot}$~\citep{gora:ephem} to 15 M$_{\odot}$~\citep{bowler:mass}. The orbital motion of the gainer is traced by the He II $4685$ \AA\ line.~\citet{fabrika:massfunction} were the first to derive a value for the mass function of the system. They obtained a semi-amplitude for the compact object of $K_X=175\pm20$ km s$^{-1}$ and a systemic velocity $\gamma=-13\pm13$ km s$^{-1}$ centered around precessional phase $\psi\sim0$, when the central part of the disk is presented nearly face-on. The He II line displays the double peak profile expected from a  thin disk, but is highly variable and the relative peak intensities change with precessional phase. The consensus is that the He II $4685$ \AA\ has contributions besides the disk (for example, from an extended region that includes the base of the jets).
 
The determination of the velocity curve of the loser is far more difficult.~\citet{gies:two} and~\citet{gies:one} succeeded in distinguishing the spectrum of the companion. They identified a spectral region with many weak absorption lines of singly ionized metals, e.~g., Fe II, that correspond to a A $4-7$ supergiant spectrum. They took the supergiant HD9233 as the reference star (the superposition of the choosen spectral region between SS 433 and HD9233 is quite similar), and used cross correlation to obtain the radial velocities. This was subsequently refined by~\citet{hillwig:one} and~\citet{hillwig:two}. They determined M$_O=12.3\pm3.3$ M$_\odot$ for the A star and M$_X=4.3\pm0.8$ M$_\odot$ for the black hole (BH)~\citep[their radial velocity curve for the compact object is consistent with][]{fabrika:massfunction}. They observed the system at $\psi\sim0$, to reduce the systematics arising from any disk flows, obtaining $K_O=58.2\pm3.1$ km s$^{-1}$ and $\gamma=73\pm2$ km s$^{-1}$, where $K_O$ and $\gamma$ are the semi-amplitude and systemic velocities, respectively.  
More recently, \citet{kubota:subaru} used high resolution spectra confirming the results of~\citet{hillwig:two}, although they included the~\citet{hillwig:two} data in their analysis. They derived the velocity curve of the compact object using four He II lines. They fix the systemic velocity obtained from the A star velocity curve, $\gamma_O = (59.2\pm2.5)$ km s$^{-1}$. 

\begin{figure}
\centering
\resizebox{\hsize}{!}{\includegraphics{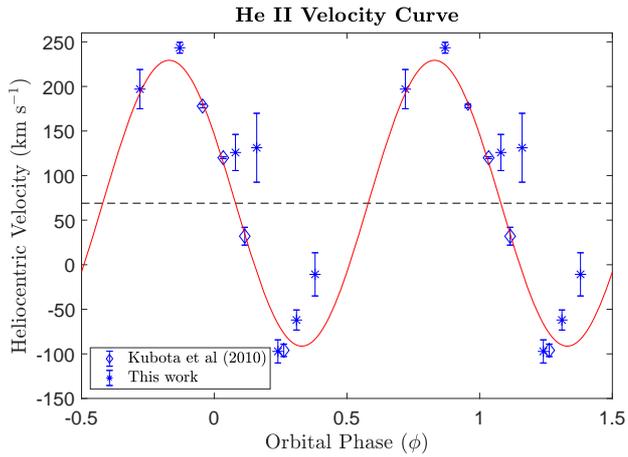}}
\caption[3pt]{Velocity curve of He II 4686\AA. The two points at $\phi\sim0.1-0.2$ are probably affected by intrinsic variations in the line forming region and are not considered in the fit.}
\label{fig:heIIcurve}
\end{figure}

We have refined the fit, adding the He II velocities from the XSHOOTER spectra at face-on presentation.  The new velocity curve,  assuming a circular orbit here and in all subsequent fits, is shown in Fig.~\ref{fig:heIIcurve}.  
The velocities were measured by calculating the line center of flux following~\citet{kubota:subaru}. The associated error depends on the line width and complexity; the line was not fitted by a profile function to avoid bias. The orbital parameters are shown in Table~\ref{table:fitcomponents}. The velocity semi-amplitude is consistent with~\citet{fabrika:massfunction}. At $\phi\sim0.1-0.2$ there are two points clearly separated from the curve that we did not include in the fit. We suggest that the line, even if mainly tracing the gainer's orbital motion, has a variable contribution related to a stream-disk impact hot spot or a wind.  During those nights no outburst was reported. We emphasize that this system can be strongly affected by transient line variations and other multiple components  can modify the line profiles, although the overall profile is more stable.  
\subsubsection{Mass, radius and rotational velocity estimation of the A star}
\label{subsec:astar}

\begin{figure}
\centering
\resizebox{\hsize}{!}{\includegraphics{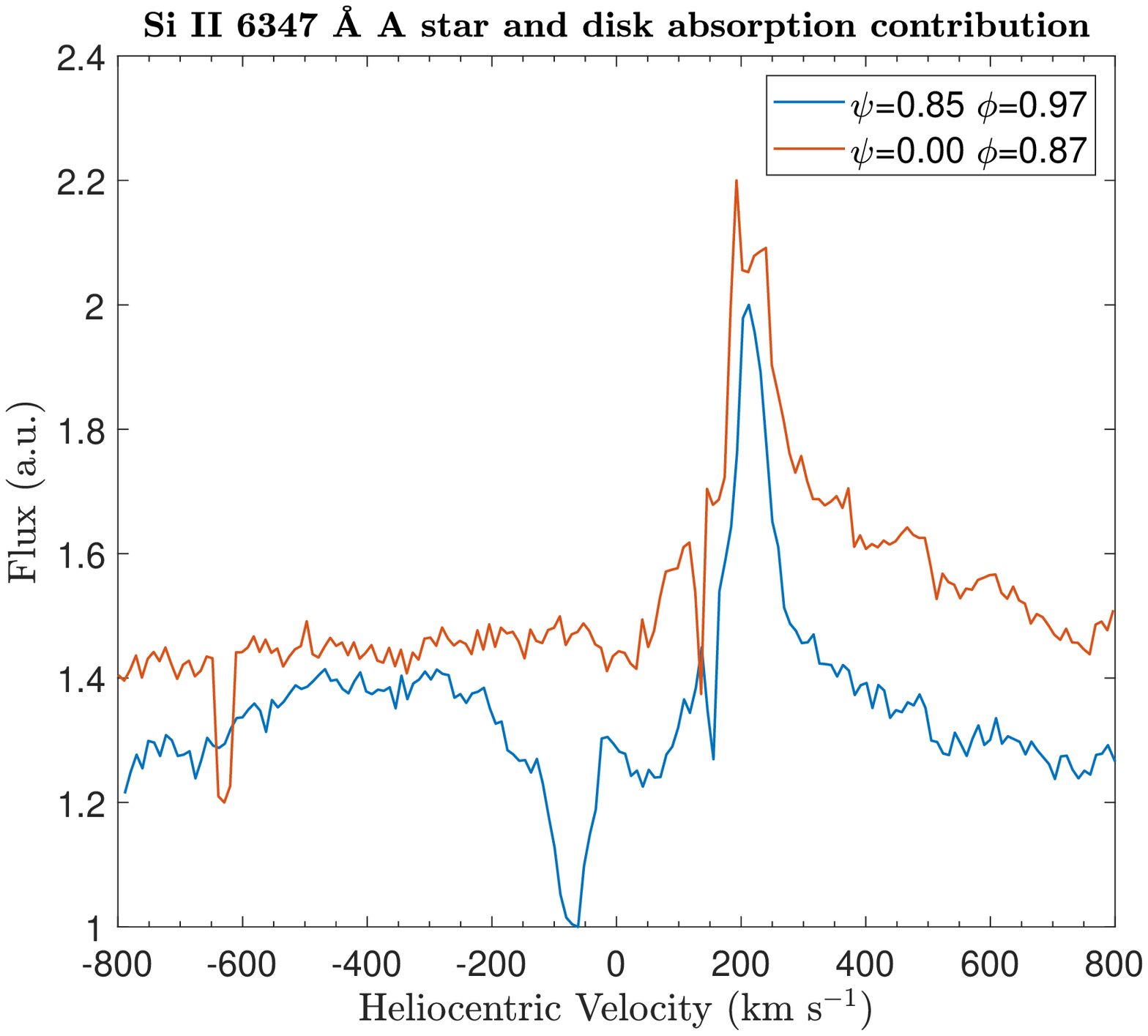}}
\caption[3pt]{Si II 6347 \AA\ absorption at two different precessional phases.  Only the A star absorption is present when the disk is face-on, while the disk also contributes at $\psi\sim0.8$.}
\label{fig:SiIIdiskStar}
\end{figure}

\begin{figure}
\centering
\resizebox{\hsize}{!}{\includegraphics{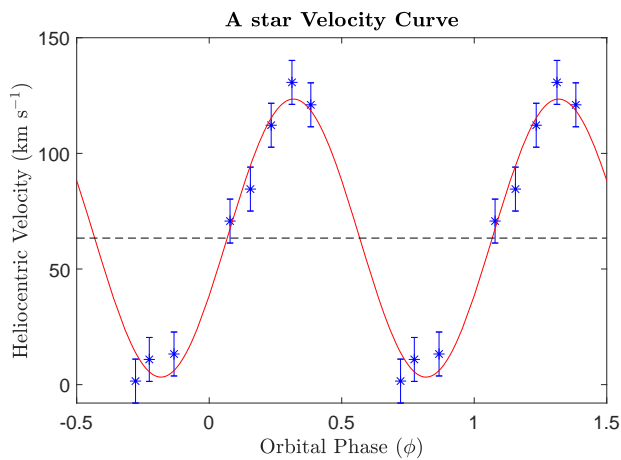}}
\caption[3pt]{Velocity curve of the A supergiant. The spectra chosen were when the disk was nearly face-on.  See text for discussion.}
\label{fig:starcurve}
\end{figure}

\begin{figure}
\centering
\resizebox{\hsize}{!}{\includegraphics{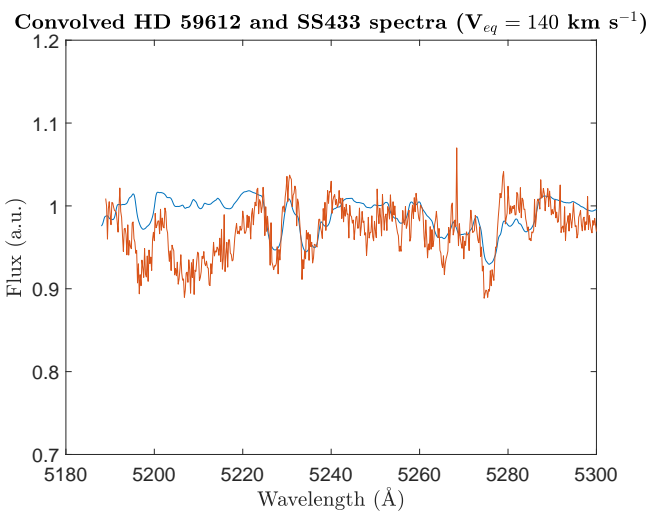}}
\caption[3pt]{Comparison between the A star standard spectrum diluted with a continuum contributing half the total light.}
\label{fig:astardisk}
\end{figure}

The Si II $6347$ \AA\ line showed weak absorption components at precessional phases $ 0.9 <\psi < 0.1$ whose  velocity changes with orbital phase. At these precessional phases, the  lines are less affected by possible contamination from the disk. We exploit this new result in our final velocity curve without using cross correlations; the result is shown in Fig.~\ref{fig:starcurve} and the  derived parameters are listed in Table~\ref{table:fitcomponents}. The values are consistent with those found by \citet{kubota:subaru, hillwig:two}. There is also agreement with the $\gamma$ velocity of the He II curve, indicating, again, that these transitions trace the two components of SS 433, whose systemic velocity is higher than $+60$ km s$^{-1}$. The A star lines are also detected at intermediate precessional phases. The Si II transitions are also present at near edge-on phases.  They are partially blended with the disk Si II lines, but these are at higher velocity, stronger intensity in absorption, and  narrower (see Fig.~\ref{fig:SiIIdiskStar}).  No velocities were measured at these phases because of possible systematics affecting the lines.
\begin{table}
\caption{Fit parameters of the binary components of SS 433. $\phi_0$ is in orbital phase units.}
\begin{tabular}{cccccc} 
\hline\hline
Line & K (km s$^{-1}$) & $\phi_0$ & $\gamma$ (km s$^{-1}$) & $\frac{\chi^2}{\nu}$ \\
\hline
He II 4685 \AA\ & 160.3$\pm$3.1 & 0.42$\pm$0.03 & $69.0\pm$4.7 & 8.76 \\
Si II 6347 \AA\ & 60.1$\pm$4.3 & -0.07$\pm$0.02  & $63.4\pm$3.5 & 0.92\\
\hline
\label{table:fitcomponents}\\
\end{tabular}
\end{table}
From the parameters  in Table~\ref{table:fitcomponents}, we derived the masses of the SS 433 system: $M_O=11.3\pm0.6$ M$_\odot$ for the supergiant and $M_X=4.2\pm0.4$ M$_\odot$ for the compact object, indicating again that it is a low mass BH.  For a circular orbit, Kepler's third law yields the binary separation $a = 0.271\pm0.004$ AU = 58.3$\pm$0.9 R$_\sun$. Using~\citet{eggleton:rochelobe} for the Roche radius
\begin{equation} 
\label{eqn:rochelobe}
\text{R}_\text{L}=\frac{0.49q^{-2/3}}{0.6q^{-2/3}+\ln(1+q^{-1/3})}a ,
\end{equation}
where $a$ and $q = \frac{M_X}{M_O}$ are, respectively, the binary separation and the mass ratio of the two components, we obtain $R_L= 27.2\pm2.6\ \text{R}_\odot$.  This is similar to~\citet{hillwig:one}.  We used the XSHOOTER spectra to estimate the radius (R$_O$) and the equatorial rotational velocity V$_{eq}$ of the A star using the A4 Ib standard HD $59612$  from the ESO archive, choosing the spectral region between $5200$ \AA\ and $5350$ \AA\ where the signal to noise of the XSHOOTER spectra was sufficiently high for the comparison, and selected the SS 433 spectrum at precessional phase $\psi=0.91$, where the disk absorption line contribution is reduced. For simplicity we assume synchronism since tidal forces tend to enforce it  on a timescale similar to that for circularization of the binary orbit~\citep{zahn:tide,frank:accretion}.  We note, however, that this is only an assumption in the absence of detailed evolutionary modeling of the progenitor system. We adopted a limb darkening parameter $u=0.5$ for the convolving profile for a rigid rotator. Fig.~\ref{fig:astardisk} shows the comparison of the SS 433 spectrum with the convolved profile of HD $59612$. We also included a constant contaminating continuum from the disk, the level of which was adjusted during the comparison.   The required value, about 50\% of the flux at these wavelengths, is similar to that inferred from the PL variations during primary eclipse (Sect.~\ref{subsec:pollight}).  We used 80 - 160 km s$^{-1}$, spanning the range in the literature, finding an acceptable match  with \textbf{V}$_{eq}=(140\pm20)$ km s$^{-1}$. Assuming synchronous rigid body rotation gives a radius is R$_O = (36\pm5)$ \text{R}$_\odot$. This is somewhat higher than the Roche Lobe value we derived, consistent with the A star filling or overfilling its Roche lobe.

\subsubsection{Disk-wind and structure of the accretion disk}
\label{subsec:diskstructure}
A blueshifted absorption, the signature of a net outflow, is seen when the disk is edge-on. This feature is persistent and varies similarly for all species on the precessional cycle. This has also been discussed by~\citet{fabrika:review} and~\citet{fabrika:wind}, who tracked the velocity of the outflow as a function of precessional phase. When the disk is nearly face-on, v$_{\text{rad}}$  is high (from $\sim-180$ to $\sim-1300$ km s$^{-1}$). 

\begin{figure}
\caption[3pt]{Velocity curve of the three Paschen lines. The diamonds are the velocities of the Pa $12$, the squares of Pa $11$, and the stars of Pa $10$.}
\label{fig:pascurves}
\centering
\resizebox{\hsize}{!}{\includegraphics{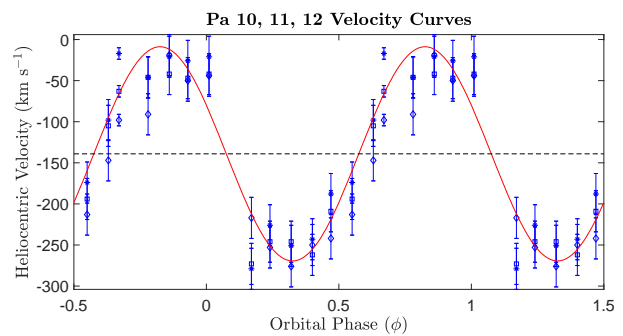}}
\end{figure}

\begin{figure}
\centering
\resizebox{\hsize}{!}{\includegraphics{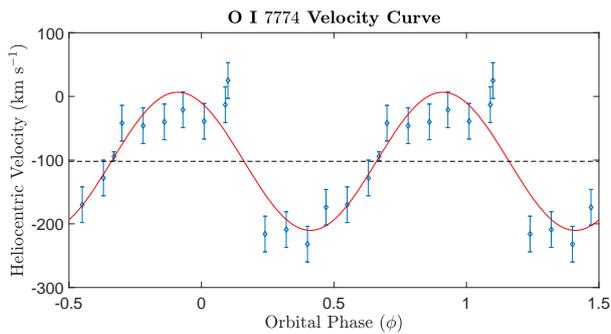}}
\caption[3pt]{Velocity curve of the O I $7774$ \AA\ line. }
\label{fig:ox7774}
\end{figure}

\begin{figure}
\centering
\resizebox{\hsize}{!}{\includegraphics{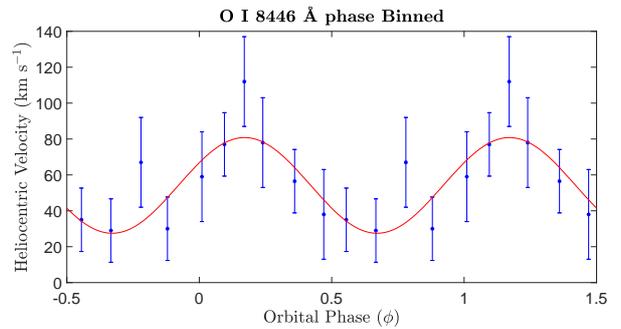}}
\caption[3pt]{Velocity curve of the O I $8446$ \AA\ line from the EMMI dataset when the disk is nearly edge-on. Some points were phase binned to improve the statistics since the resolution of the spectrographs is low (25 km s$^{-1}$ px$^{-1}$). This line is discussed in Sects.~\ref{subsec:diskstructure},~\ref{subsubsec:wind} and~\ref{subsec:explanation}.}
\label{fig:o8446_binned}
\end{figure}

\begin{figure}
\centering
\resizebox{\hsize}{!}{\includegraphics{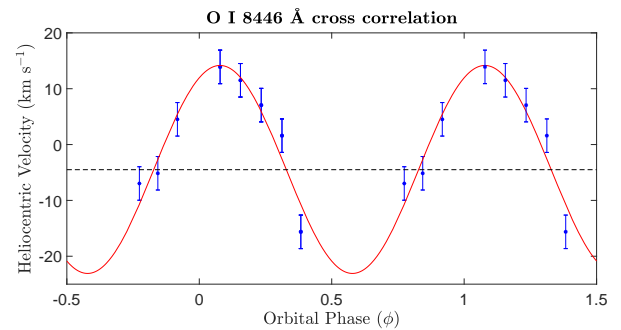}}
\caption[3pt]{Velocity curve of the O I $8446$ \AA\ line from XSHOOTER. The velocities are from the June dataset at nearly phase-on precessional phase and were  cross correlated with the point at $\phi=0.08$. Superposing the $\gamma$ velocity of the system (from the He II or Si II curve for example) gives a value in agreement with Fig.~\ref{fig:o8446_binned} and with the system velocity, indicating that this line does not come from the disk-wind.}
\label{fig:o8446_cross}
\end{figure}

\begin{figure}
\centering
\resizebox{\hsize}{!}{\includegraphics{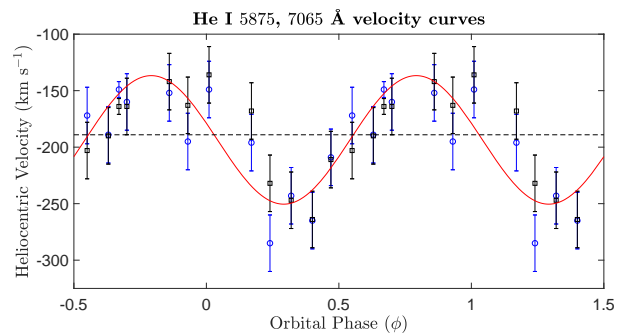}}
\caption[3pt]{Velocity curves of the He I Lines. The red line represents the global fit. The circles are the He I 7065 \AA\ line while the black squares the He I 5875 \AA.}
\label{fig:hecurves}
\end{figure}

To search for  orbital modulation of the absorption velocity, we chose the interval $0.2 \le \psi \le 0.4$ where our phase coverage is good. The velocity curves were constructed from the EMMI spectra (we used only one XSHOOTER spectrum at the complementary phase $\psi=0.66$) for all lines that show the absorption, that is,  O I $7774$ \AA, O I $8446$ \AA, Pa $10$-$12$, He I $7065$ \AA, He I $6678$ \AA\ and He I $5875$ \AA. We measured the  velocity of the deep absorption, using the EMMI velocity resolution (25 km s$^{-1}$) to estimate the uncertainty.  All  curves have been obtained near edge-on presentation of the disk except Fig.~\ref{fig:o8446_binned}, at near face-on phases. These lines show clear orbital modulation with different amplitudes depending on the line formation regions. The curves are shown in Figs.~\ref{fig:pascurves}--\ref{fig:hecurves}. 
All of the radial velocity curves are synchronous in phase, apart from the O I 8446 \AA, and vary with the same orbital period as He II 4686 \AA. The Paschen velocities are shown together in Fig.~\ref{fig:pascurves} to manifest their close correspondence. Table~\ref{table:pasfit} lists the fit parameters for each line. The aggregated Paschen line fit gives K=130.3$\pm$5.2 km s$^{-1}$, $\phi=0.43\pm$0.01 and $\gamma=-139.1\pm$3.8 km s$^{-1}$ with a $\chi^2/\nu=3.88$. 
\begin{table}
\caption{Fit parameters of all the lines considered. The two O I 8446 \AA\ transition  are from EMMI and cross-correlated XSHOOTER analyses, respectively. (Figs.~\ref{fig:o8446_binned},~\ref{fig:o8446_cross}).  $\phi_0$ is in orbital phase units.}
\begin{tabular}{ccccc} 
\hline\hline
\textbf{Line} & K (km s$^{-1}$) & $\phi_0$ & $\gamma$ (km s$^{-1}$) & $\frac{\chi^2}{\nu}$ \\
\hline
Pa 10 & 125.1$\pm$9.6 & 0.38$\pm$0.01 & $-149.1\pm$7.4 & 1.40 \\
Pa 11 & 129.0$\pm$9.7 & 0.38$\pm$0.01  & $-137.5\pm$7.4 & 2.01\\
Pa 12 & 123.8$\pm$9.5 & 0.39$\pm$0.01  & $-151.1\pm$6.6 & 0.67\\
\\
He 5875 \AA\ & 53.4$\pm$10.0 & 0.38$\pm$0.02 & $-184.5\pm$6.6 & 0.67 \\
He 7065 \AA\ & 56.9$\pm$8.7 & 0.46$\pm$0.03  & $-193.6\pm$6.6 & 1.38\\
He 6678 \AA\ & 105.1$\pm$12.6 & 0.39$\pm$0.02  & $-134.9\pm$9.5 & 1.30\\
\\
O 7774 \AA\ & 108.4$\pm$10.9 & 0.34$\pm$0.01  & $-101.9\pm$6.7 & 2.23\\
\\
O 8446 \AA\ & 18.6$\pm$2.2 & 0.17$\pm$0.01 & $-4.5\pm$1.3\tablefootmark{a} & 1.4\\
		& 26.7$\pm$8.9 & 0.08$\pm$0.06 & 54.1$\pm$6.5\tablefootmark{b} & 0.7\\

\hline
\label{table:pasfit}\\
\end{tabular}
\tablefoot{
\tablefoottext{a}{Fit parameters obtained from the velocities shown in Fig.~\ref{fig:o8446_cross}. All the velocities have been heliocentric corrected only. }
\tablefoottext{b}{Fit parameters obtained from the velocities shown in Fig.~\ref{fig:o8446_binned}. All the velocities have been obtained by measuring the deep absorption in the core of the line.}
}
\end{table}

The Paschen lines, formed by recombination, show that the absorption when the disk is edge-on is due to self absorption by the disk and the wind. In contrast, the He I and O I velocity curves show different velocity amplitudes for lines of the same ion; different lines trace different regions.
\begin{figure}
\resizebox{\hsize}{!}{\includegraphics{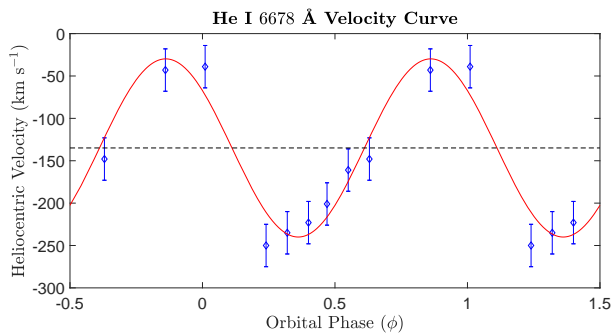}}
\caption[3pt]{Velocity curve of the He I $6678$ \AA. This curve has fewer points because some He I profiles were blended with the redshifted H$\alpha$ jet line.}
\label{fig:he6678curve}
\end{figure} 
The intercomparison of the velocity curves reveals a correlation of the systemic velocity, $\gamma$, with the velocity semi-amplitude, i.~e., curves with similar velocity amplitudes show the same $\gamma$. Thus, the local environment sampled by each transition has a velocity component which is not purely orbital in the frame of the binary, or is coming from a different plane than the binary motion.

\begin{figure*}
\centering
   \includegraphics[width=17cm]{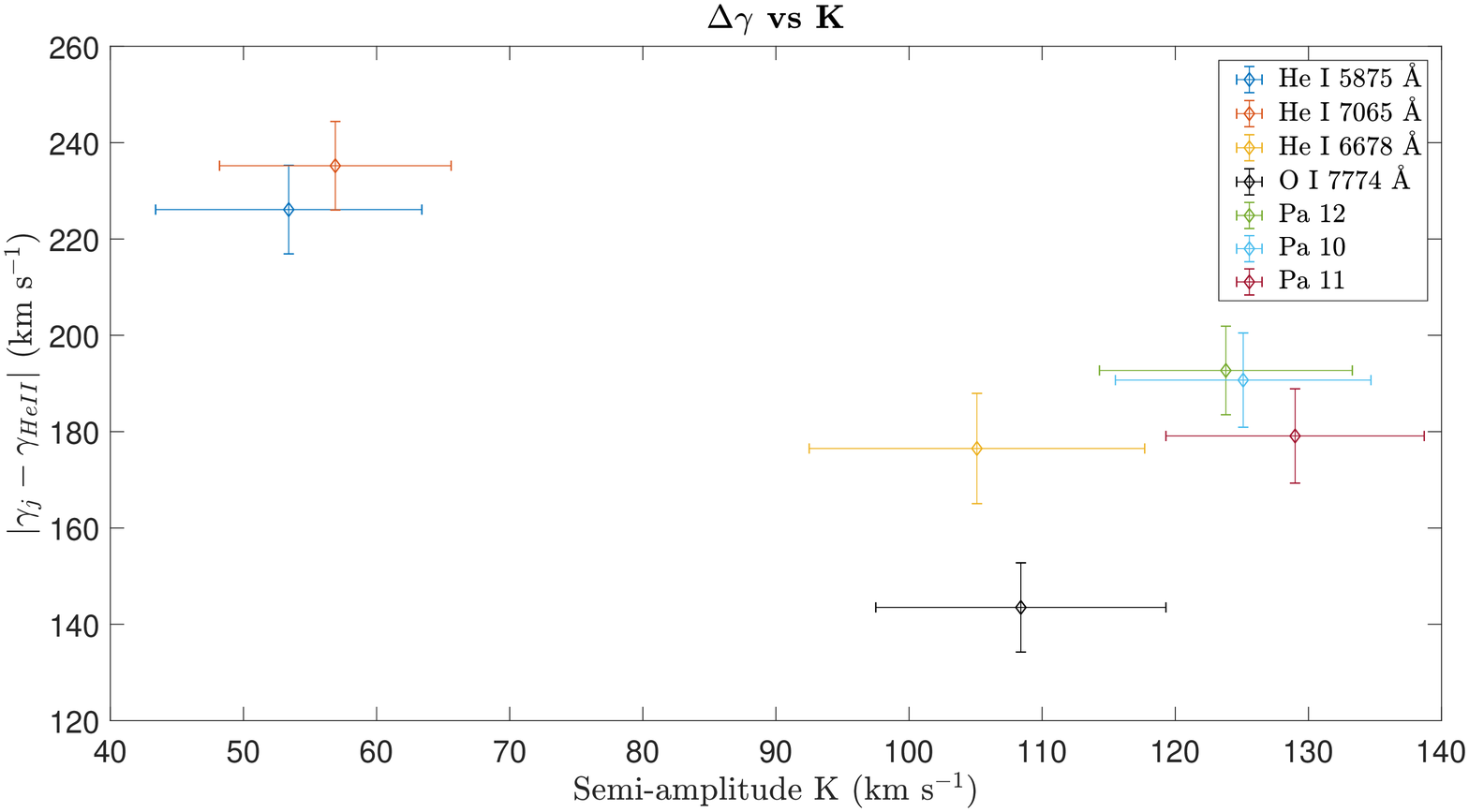}  
     \caption{Systemic velocities $\gamma_j$ minus the system velocity (taken as that of He II) as function of the semi-amplitude velocities K.}
     \label{fig:veldist}
\end{figure*}
\subsubsection{Wind velocity}
\label{subsubsec:wind}
The velocity curves indicate an outflow from the disk. From the precessional modulation of the absorption (Sect.~\ref{sec:analysis}) we infer that this structure is geometrically thick. The velocity of each transition at the optically thick surface, the Pseudo-photosphere (hereafter \textit{Pp}), is given by $\Delta\gamma$, obtained from the difference between the individual velocity curves and the true systemic velocity from the He II curve. The velocities measured possess a component from the binary motion, and a component from the wind  that is independent of orbital phase $\phi$ and depends only on the distance to the gainer. So, since every line traces a region whose distance from the gainer is fixed, $\Delta\gamma$ should indicate the wind velocity at the line $\tau\simeq1$ surface for each transition. The wind velocities are shown in Fig.~\ref{fig:veldist}.

Fig.~\ref{fig:veldist} shows that  He I 6678 \AA\ singlet forms in a region where the wind gradient is higher, near the plane of the disk, and near the inner regions of the disk.  In contrast,  the  He I triplet lines trace the outer part of the disk wind in height and radial distance. The O I $7774$ \AA\ and He I $6678$ \AA\ have similar intrinsic strengths ($\log gf$) and trace the same part of the wind.  The observed Paschen lines, from high members of the series, have similar velocities as  He I $6678$ \AA; their optical depth is much lower than the principal Balmer lines. 

Finally, Figs.~\ref{fig:ox7774},~\ref{fig:o8446_binned} and~\ref{fig:o8446_cross} show that the O I $8446$ \AA\ and $7774$ \AA\ transitions sample different regions. The O I $7774$ \AA\ traces a structure that is circulating and outflowing, like the other transitions. The O I 8446 \AA\ has a systemic velocity consistent with He II 4686 \AA, while the 7774 \AA\ multiplet shows the same shift as the Paschen lines.   The O I absorption lines show that the disk and outflow self-absorb along the line of sight. The face-on systemic velocity is that of the BH indicating that the O I  7774 \AA\ is optically thin for out-of-plane flow.  Seen face-on, $\gamma$ for O I 8446 \AA\ again agrees with He II 4686\AA.

\begin{figure}
\centering
\resizebox{\hsize}{!}{\includegraphics{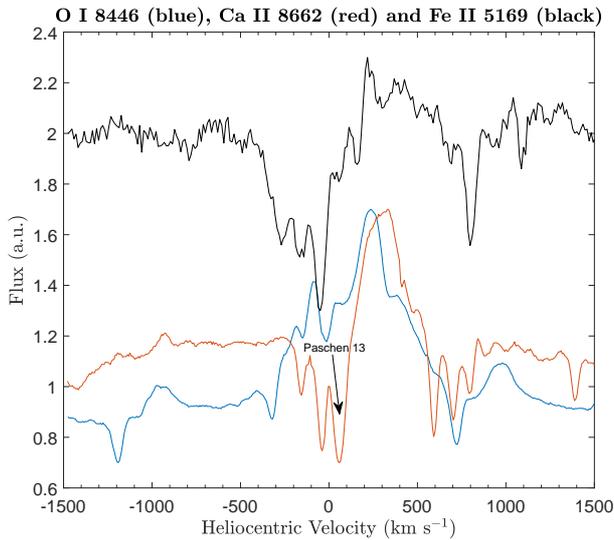}}
\caption[3pt]{O I $8446$ \AA, Ca II $8662$ \AA\ and Fe II 5169 \AA\ transitions. The absorption features match in velocity. The Paschen 13 absorption feature is indicated by the arrow. The phase of the displayed spectrum is $\psi=0.87$ and $\phi=0.10$.}
\label{fig:ocaii}
\end{figure}

The phasing of the radial velocity for O I 7774 \AA\ vs 8446 \AA\ points to the latter arising mainly at the L1 point on the A star. Its presence as an emission line can be explained by Ly$\beta$ fluorescent excitation of the resonance transition at 1025 \AA\ \citep[e.g.,][]{keenan:oi}, on exposure of the substellar point to the accretion disk~\citep[a Grotrian diagram is shown in][]{shore:oi}. This is the only line in the optical spectrum with these particular properties. The phase shift indicates multiple contributors with radial velocity variations, that is,  the disk being in antiphase with the L1 point, but the relative amplitudes possibly changing over time. Another transition that forms because of radiative pumping is the Ca II 8662 \AA\ Infrared Triplet (IRT). It is partially blended with Pa 13 and is very weak, but it is well distinguished in the XSHOOTER spectra taken between $\psi=0.86-0.89$. Both transitions are shown Fig.~\ref{fig:ocaii}. The IRT show two absorption components at $\sim-150$ and $\sim-40$ km s$^{-1}$, while Pa 13 shows only the $\sim-40$ km s$^{-1}$ feature. If the O I 8446 \AA\ semi-amplitude is from orbital motion of L1, then it provides another estimate of the Roche radius, independent of the rotational velocity. Using the same argument as above, R$_{RL} \approx 0.2a$, which is consistent with that obtained from both the rotational analysis and the mass ratio for R$_{RL}$. Thus, the combination of emission from material falling toward the gainer and being illuminated on the sub-stellar face of the loser appears to account for the observed O I 8446 \AA\ variations. 
There is no need to invoke an orbiting circumsystem disk to account for the velocity and profile variations.  Evidence based on profile decomposition as a set of independent contributors can also be affected in lower resolution spectra by intrinsic line profile variations that occur on timescales as short as a single period,  such as the H$\alpha$ variations reported by~\citet{blundell:circ}.  Although a portion of the data set we used are the same spectra as~\citet{blundell:circ}, we reach a different conclusion when those are supplemented by higher resolution data.  Further dynamical evidence for a circumbinary disk has been adduced from a similar decomposition analysis of the O I profile~\citep[e.g., ][]{bowler:circ2010, bowler:circ2011, bowler:circ2013}. The higher resolution data we have used in this study do not support this.  Indeed, much of the variation in the relative peak intensities is due to variable absorption (e.g., Fig.~\ref{fig:ocaii}) that have been treated as independent components in the published  analyses.

\subsubsection{Estimate of the disk radius}
\label{subsec:diskradius}
An estimate of the disk radius is provided by the locale where the  specific circular angular momentum is the same as that transferred through the L$_1$~\citep{shu:review, weiland:hotspot}. The specific angular momentum of the disk is $j_{disk}=(GM_\mathrm{x}r_\mathrm{disk})^{\frac{1}{2}}$, where $M_\mathrm{x}$ is the mass of the gainer, and that carried by the stream is  $j_{stream}=(1-R_{RL}/a)^2a^2\Omega$  where $\Omega$ is the orbital frequency of the binary  and $a$ the separation of the components. Equating these gives $r_{disk}/a=0.30$ for $q=0.37$, that is, r$_{disk}$ = 17 R$_\odot$. In the end, $ 17 \le r_{disk}\le 31$ R$_\odot$, the upper limit coming from the binary separation and the A star Roche Lobe radius.
\subsubsection{Multiple absorption components}
\label{subsec:afs}

An XSHOOTER spectrum from 2017 June 30.2 ($\psi=0.91$, $\phi=0.72$), showed complex line profiles (the He I 7065 \AA\  is shown in Fig.~\ref{fig:heprecmod1}). The blueshifted absorption components at low and high velocity (e.~g., He I $7065$ \AA, O I $7774$ \AA) suggest an intense mass ejection from the wind, revealed by mass loading of the outer parts of the outflow.  Some lines, especially those from species with low ionization potentials (e.g.,  Fe$^+$, C$^+$, and Ca$^+$) show the absorption only at high velocity, indicating that they are tracing the outer wind. If the picture outlined in Sect.~\ref{subsubsec:wind} is correct then when the disk is nearly face-on, the full wind gradient is sampled by the line of sight.  If the system undergoes an increased mass ejection, it may be possible to see the absorptions at low velocity even when the disk is face-on.  We note that this event was not related to a flare of SS 433.  That these properties are not unique is demonstrated by a second XSHOOTER sequence obtained on 2018 May 12.2 -  17.2 ($0.86\le \psi \le 0.89$, $0.94 \le \phi \le 0.25$); see Fig.~\ref{fig:ocaii} as an example.  Again, no flare was reported during this sequence.  For H$\beta$, H$\gamma$, and H$\delta$, the absorption extends to $-450$ km s$^{-1}$ but the profile shows a sharp velocity maximum in the last spectrum. The redward emission wing extends to +800 km s$^{-1}$ throughout the sequence.  The emission component of the profile remains constant for the first four spectra but halves in the last spectrum. In the last spectrum of the 2018 sequence, the He I 5876 \AA\ and 7065 \AA\ lines also show enhanced absorption to $-400$ km s$^{-1}$ and a persistent low velocity feature at $-110$ km s$^{-1}$; in contrast,  He I 6678 \AA\  shows only the low velocity component.  The Fe II lines 4921, 5018, and 5169 \AA\ all show the same profiles as the He I triplets but with multiple components ( $-300$, $-200$, $-80$ km s$^{-1}$); only the lower two persist. The absorption profile is strongly phase dependent. Similar line profiles were also observed by~\citet{robinson:HK} (see bottom panel of Fig. 2). The appearance of multiple, low dispersion absorptions points to an inhomogeneous and clumpy wind.
\subsubsection{Spectra during the 2018 outburst}
\label{sec:outburst}
The LT sequence was fortuitously obtained during a radio outburst of SS 433, detected by~\citet{gora:atel}, it started on $2018$ Jul $17.772$ and ended on Jul $18.022$ UT. At the same time, an increase in the optical brightness was detected up to $R = 11.7$, an increase of $\sim2$ magnitudes above the mean value. The LT covered the last part of this burst.  The photometry shows an increase in brightness for the orbital phase at which the burst started.  \citet{gora:atel}  reported  increased emission only in He II 4686 \AA.  The LT spectrum, that covered only the red spectral interval, did not show any particular variation.  The phase of the burst coincided with those of~\citet{kopylov:burst} and~\citet{blundell:burst}. This coincidence between orbital phase and flaring activity may be pointing to an orbital ellipticity, as is known from other microquasars (e.~g. LS+$61^\circ 303$).

Polarimetric LT observations were also obtained for this night (black points from Figs.~\ref{fig:sblightcurve} to~\ref{fig:srlightcurve}). No dramatic change was detected in either PL or PA during the burst night in any of the three filters.

\begin{figure}
\centering
\resizebox{\hsize}{!}{\includegraphics{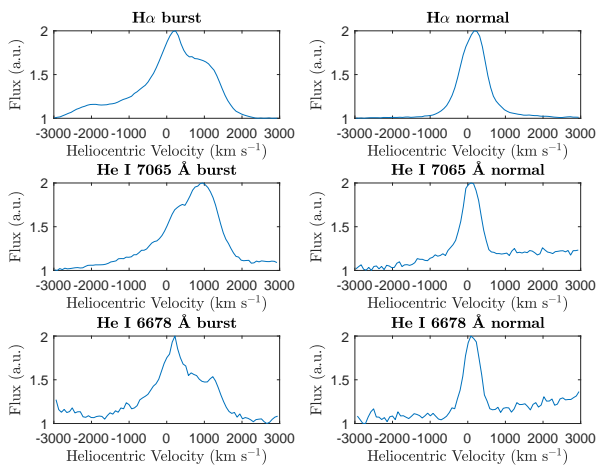}}
\caption[3pt]{Comparison of H$\alpha$, He I $7065$ \AA\ and He I $6678$ \AA\ before and during the burst.}
\label{fig:burstlinecomparison}
\end{figure}

The spectra show no particular changes until $23$ days after the burst, corresponding to nearly two orbits, when three lines show far higher intensities and increased widths\footnote{The H$\alpha$ line shows no intensity increase but it forms over a very spatially extended region and may be less sensitive to the more local changes induced in the disk following a burst. Unlike the He I lines, there were other nights for which H$\alpha$ showed an enhanced peak to continuum ratio.}. Fig.~\ref{fig:burstlinecomparison} shows the comparison of H$\alpha$, He I $7065$ \AA\ and He I $6678$ \AA.
The lines are very broad and display multiple features at high velocities, around $\pm1000$ km s$^{-1}$. In particular, H$\alpha$  shows a double redshifted structure which persists for the whole observational interval. The neutral helium lines, instead, show the anomalous profile only on Aug 1.004 ($\psi=0.35$, $\phi=0.04$).  The subsequent spectra were taken six days later, so it is not possible to know the time scale over which they returned to quiescence. There are notable similarities in Fig.~\ref{fig:burstlinecomparison} between the three profiles during the burst, so the changes caused by the burst appear to be from a different region than that responsible for the quiescent lines.

The common redshifted emission, $+1000$ km s$^{-1}$  and the blue wing  suggest that these  lines, even if intrinsically different from each other and coming from different parts of the equatorial structure (see Sect.~\ref{subsec:diskstructure}), show the same dynamical variation, perhaps the formation of outward propagating optically thin bright spots. 

A comprehensive follow-up study of a radio burst of SS 433 was presented by~\citet{blundell:burst} using the same EMMI spectroscopic data set of~\citet{blundell:circ} that we used in this work. Beginning at a phase where the spectra were normal, H$\alpha$ started to show the extended redshifted wings. They observed that after five days the jet lines intensified and their velocities ($\beta\sim0.28-0.29$) increased. Three days later, a type 1 radio flare occurred.   The new LT spectra agree with the flaring line profile behavior described by~\citet{blundell:burst}.  \citet{gora:atel} found  enhanced He II line the same day of the burst. It is the only line that shows an instantaneous response to the flare, meaning that the flare comes from the inner parts of the accretion disk. The fact that H$\alpha$ and, most important, the He I lines show a change in the profiles days after the burst is indicative of a delayed response of the accretion disk to the flare. From the velocity curves derived in this work, moreover, the line forming regions are known for He I $7065$ \AA\ and He I $6678$ \AA\ so it appears that the profile changes are from the disk-wind. The constancy of PL and PA, on the other hand, indicates that the outburst does not affect the scattering structure.

\section{Conclusions}
\label{subsec:explanation}

Concentrating on the spectral sequence from the near edge-on presentation of the disk permits a separation of the circulatory and outflow motions of the disk gas.  The outflow, indicated by the P Cyg lines, appears tied to the orbital motion of the gainer.  Choosing those transitions, that are sufficiently optically thick, here labelled $j$,  that we can locate the \textit{Pp} from the absorption edges, we measured their orbitally modulated radial velocity amplitude, $K_j$ and systemic velocity, $\gamma_j$. We assume as symmetric outflow from the gainer as the basic geometry, whether axial or spherical is not important. Then the difference between $\gamma_{He II}$ and $\gamma_j$ parameterizes the outflow at the {\it Pp} and the amplitude $K_j$ gives the distance of that surface from the binary center of mass. We note that for all lines showing a strong P Cyg absorption component, $\gamma_j < 0$. Although non-axisymmetric structures complicate this picture, as do the stream and hot spot,  this should not compromise the mapping of the velocity for the low ionization species we used. The outflow is ionization stratified. Since the central engine is the ionizing source, we would expect that those lines tracing progressively higher ionization states will have smaller velocity amplitudes. The O I transitions are especially important in tracing  this because neutral oxygen is partially shielded by hydrogen; consequently,  O I lines should  be formed in the peripheral regions of the disk but within the region traced by the ionized metals. The Paschen lines are formed by recombination but those we have available are high in the series and weaker than the principal Balmer lines, so they probe deeper portions of the outflow. Further information is provided by the orbital modulation as a function of phase.  When seen nearly face-on, the He I lines, for instance, do not show a P Cyg absorption trough and have greater widths but still vary in radial velocity.  Together, this suggests a  flaring wind of the sort modeled by~\citet{proga:wind2} in which there is an acceleration and outflow from above the disk in a sort of biconical geometry. The weaker lines, especially  Fe II and Si II, also probe deeper portions of the wind.  The derived velocity gradient, shown in Fig.~\ref{fig:veldist}, is consistent with a slow acceleration of the outflow and will help constrain models of the line formation.  We leave that for future work. We cannot derive a mass loss rate directly from these measurements without a better handle on the opening angle of the flow, but the dynamics are consistent with a slow, weakly collimated wind like those from the  hydrodynamic simulations.  A wind also accounts for the narrow absorption components. These are confined within the velocity range of the broader P Cyg absorption trough but vary sporadically over time. The integrated strengths of the disk lines vary, as well so these features may be like the discrete absorption features seen in hot star winds. With our limited coverage over the long term, however,  we cannot  determine whether a thermal wind or radiative acceleration is responsible for the driving.

\begin{acknowledgements}
We thank Vilppu Piirola for the Dipol-2 and Dipol-UF observations. P.P. and S.N.S. thank Linda Schmidtobreick for generously furnishing the EMMI spectra and Anatol Cherepashchuk for correspondence. P. P. thanks Ferdinando Patat for crucial information on polarimetry data reduction and analysis given at the polarimetry school in Padova ``Looking at Cosmic Sources in Polarized Light''. We acknowledge support from the ERC Advanced Grant HotMol ERC-2011-AdG-291659. The Dipol-2 and Dipol-UF polarimeters were built in cooperation by the University of Turku, Finland, and the Leibniz Institut f{\"u}r Sonnenphysik, Germany, with support from the Leibniz Association grant SAW-2011-KIS-7. We are grateful to the Institute for Astronomy, University of Hawaii, for the allocated observing time. Based on observations made with the Nordic Optical Telescope, operated by the Nordic Optical Telescope Scientific Association at the Observatorio del Roque de los Muchachos, La Palma, Spain, of the Instituto de Astrofisica de Canarias. Based on data obtained from the ESO Science Archive Facility. We thank the referee for insightful questions, particularly those focused on the polarization results.
\end{acknowledgements}

\bibliographystyle{aa} 
\bibliography{bibliography} 

\begin{appendix}
\section{Journal of Observations}
\label{app:journal}
Tables~\ref{table:XSHOOTER} and~\ref{table:frodospec} respectively show the journal of observations of XSHOOTER and FRODOSpec spectrographs. For the  EMMI dataset, see~\citet{blundell:burst}, as mentioned in Sect.~\ref{sec:observations}. The journal of polarization data is shown in Sect.~\ref{app:data}. 
\begin{table}[!h]
\caption{Journal of the spectra obtained from the XSHOOTER spectrograph. The observations were simultaneous in the blue and red arm of the instrument.}
\centering
\label{table:XSHOOTER}
\begin{tabular}{cccc}
\hline\hline
Calendar date & JD$-2457800$ & $\psi$ & $\phi$ \\
\hline
2017-May-21&094.78 & 0.66 & 0.67\\ 2017-May-28&101.78 & 0.71 & 0.21\\ 2017-Jun-20&124.84 & 0.85 & 0.97\\ 2017-Jun-30&134.66 & 0.91 & 0.72\\ 2017-Jul-15&149.64 & 0.00 & 0.87 \\ 2018-May-13 & 451.74 & 0.86 & 0.94\\ 2018-May-14&452.75&0.87&0.02\\2018-May-15&453.71&0.87&0.10\\2018-May-16&454.72&0.88&0.17\\2018-May-17&455.72&0.89&0.25\\ 2018-Jun-10& 479.65&0.03 & 0.08\\ 2018-Jun-11& 480.65&0.04 & 0.16 \\ 2018-Jun-12& 481.68&0.05 & 0.23 \\2018-Jun-13& 482.71& 0.05 & 0.31 \\ 2018-Jun-14& 483.64&0.06 & 0.38 \\ 2018-Jun-19& 488.74&0.09 & 0.77 \\ 2018-Jun-20 & 489.66&0.10 & 0.84\\ 2018-Jun-21 & 490.62&0.10 & 0.92\\ 2018-Jul-13& 512.60&0.24 & 0.60 \\2018-Jul-14 & 513.63& 0.24 & 0.68\\ 2018-Aug-07& 537.62&0.39 & 0.51 \\ 2018-Aug-08& 538.52&0.40 & 0.58 \\
\hline
\end{tabular}
\end{table}
\begin{table}[!h]
\caption{Journal of spectroscopic  observations from the Liverpool telescope; some nights overlap with the XSHOOTER observations.}
\centering
\label{table:frodospec}
\begin{tabular}{cccc}
\hline\hline
Calendar date&JD$-2457800$ & $\psi$& $\phi$ \\
\hline
2018-Jul-13&512.52 & 0.24 & 0.59\\ 2018-Jul-17&516.57 & 0.26 & 0.90\\ 2018-Jul-19&518.50 & 0.27 & 0.05\\ 2018-Jul-22&521.50 & 0.29 & 0.28\\ 2018-Jul-23&522.52 & 0.30 & 0.35\\ 2018-Aug-01&531.50 & 0.35 & 0.04\\ 2018-Aug-05&536.46 & 0.38 & 0.42\\ 2018-Aug-06&537.47 & 0.39 & 0.50\\ 2018-Aug-07&538.45 & 0.40 & 0.57\\ 2018-Aug-09&539.52 & 0.40 & 0.65\\ 2018-Aug-11&542.49 & 0.42 & 0.88 \\
\hline
\end{tabular}
\end{table}
\begin{table}
\caption{Journal of polarimetric  observations, Dipol-2. The second row is the outburst night.}
\centering
\label{tab:poljournal}
\begin{tabular}{cccc} 
\hline\hline
Calendar date &JD$-2457800$ & $\psi$& $\phi$ \\
\hline
2018-Jul-16&516.04 & 0.26 & 0.86\\ 2018-Jul-17&517.05 & 0.26 & 0.94\\2018-Jul-19& 519.10 & 0.28 & 0.09\\ 2018-Jul-22&522.05 & 0.29 & 0.32\\ 2018-Jul-23&523.05 & 0.30 & 0.39\\ 2018-Jul-26&526.03 & 0.32 & 0.62\\ 2018-Jul-29&529.03 & 0.34 & 0.85\\ 2018-Aug-03&534.02 & 0.37 & 0.23\\ 2018-Aug-05&536.01 & 0.38 & 0.39\\2018-Aug-06& 536.98 & 0.39 & 0.46\\ 2018-Aug-07&538.00 & 0.39 & 0.54\\ 2018-Aug-08&539.01 & 0.40 & 0.61\\ 2018-Aug-11&541.98 & 0.42 & 0.84\\ 2018-Aug-12&542.99 & 0.42 & 0.92\\ 2018-Aug-13&543.98 & 0.43 & 1.00\\ 2018-Aug-14&544.96 & 0.44 & 0.07\\ 2018-Aug-17&547.98&0.45&0.30\\2018-Aug-18&548.95 & 0.46 & 0.38\\ 2018-Aug-21&551.95 & 0.461 & 0.375\\ 
\hline
\end{tabular}
\end{table}
\section{Complementary data: Polarization and radial velocities of He II 4686 \AA, Si II 6347 \AA\ }
\label{app:data}
The complete dataset of the polarization data and the radial velocities of the He II and Si II curves (Figs.~\ref{fig:heIIcurve},~\ref{fig:starcurve}) are shown in this section. Only the measured data are shown to avoid to weigh down the text. The second row of Tables~\ref{table:buncorr} -- \ref{table:runcorr} contains the data from the burst night.
\begin{table}
\caption{Measured radial velocities of the He II 4686 \AA\ transition. All the velocities have been obtained from the XSHOOTER spectra. The error on the velocities is the resolution per pixel of the instrument. We estimate the error of the derived radial velocities by changing the limits of the wings of the profile used in the calculation of the center of emission.}
\label{table:vradheII}
\centering
\begin{tabular}{ccc} 
\hline\hline
$\psi$ & $\phi$ & V$_\text{rad}$ (km s$^{-1}$)\\
\hline
0.03&0.08&125.9$\pm$20.3\\0.04&0.16&131.2$\pm$38.6\\0.05&0.24&$-97.2\pm$13.0\\0.05&0.31&$-62.0\pm$11.3\\0.06&0.38&$-10.8\pm$24.2\\0.91&0.72&197.0$\pm$22.0\\0.00&0.87&243.4$\pm6.1$\\
\hline
\end{tabular}
\end{table}
\begin{table}
\caption{Measured radial velocities of the Si II 6347 \AA\ transition. All the velocities have been obtained from the XSHOOTER spectra. The error on the velocities is the resolution per pixel of the instrument.}
\label{table:vradSiII}
\centering
\begin{tabular}{ccccccc} 
\hline\hline
$\psi$ & $\phi$ & V$_\text{rad}$ (km s$^{-1}$)\\
\hline
0.03&0.08&71$\pm$10\\0.04&0.16&85$\pm$10\\0.05&0.23&112$\pm$10\\0.05&0.31&131$\pm$10\\0.06&0.38&121$\pm$10\\0.91&0.72&1.5$\pm$10\\0.09&0.77&11$\pm$10\\0.00&0.87&13$\pm$10\\
\hline
\end{tabular}
\end{table}
\begin{table}
\caption{Measured radial velocities of the Pa 10, 11, 12 transitions. All the velocities have been obtained from the EMMI spectra (except the XSHOOTER spectrum at $\psi\sim0.6$). The error on the velocities is the resolution per pixel of the instrument (25 km s$^{-1}$).}
\label{table:vradPa}
\centering
\begin{tabular}{cccc} 
\hline\hline
 Line & $\psi$ & $\phi$ & V$_\text{rad}$ (km s$^{-1}$)\\
\hline
Pa 10&0.31&0.01&$-21\pm$25\\&0.33&0.17&$-279\pm$25\\&0.33&0.24&$-226\pm$25\\&0.34&0.32&$-251\pm$25\\&0.34&0.40&$-243\pm$25\\&0.35&0.47&$-188\pm$25\\&0.36&0.55&$-174\pm$25\\&0.36&0.63&$-98\pm$25\\&0.66&0.67&$-17\pm$7\\&0.30&0.78&$-46\pm$25\\&0.38&0.86&$-21\pm$25\\&0.39&0.93&$-26\pm$25\\
\\
Pa 11&0.31&0.01&$-42\pm$25\\&0.33&0.17&$-273\pm$25\\&0.33&0.24&$-246\pm$25\\&0.34&0.32&$-246\pm$25\\&0.34&0.40&$-262\pm$25\\&0.35&0.47&$-209\pm$25\\&0.36&0.55&$-194\pm$25\\&0.36&0.63&$-105\pm$25\\&0.66&0.67&$-63\pm$7\\&0.30&0.78&$-46\pm$25\\&0.38&0.86&$-42\pm$25\\&0.39&0.93&$-47\pm$25\\
\\
Pa 12&0.31&0.01&$-44\pm$25\\&0.33&0.17&$-217\pm$25\\&0.33&0.24&$-253\pm$25\\&0.34&0.32&$-276\pm$25\\&0.34&0.40&$-250\pm$25\\&0.35&0.47&$-242\pm$25\\&0.36&0.55&$-213\pm$25\\&0.36&0.63&$-147\pm$25\\&0.66&0.67&$-98\pm$7\\&0.30&0.78&$-91\pm$25\\&0.38&0.86&$-19\pm$25\\&0.39&0.93&$-50\pm$25\\
\hline
\end{tabular}
\end{table}
\begin{table}
\caption{Measured radial velocities of the  He I 7065, 5875, 6678 \AA\ transitions. All the velocities have been obtained from the EMMI spectra (except the XSHOOTER spectrum at $\psi\sim0.6$). The error on the velocities is the resolution per pixel of the instrument (25 km s$^{-1}$).}
\label{table:vradHeI}
\centering
\begin{tabular}{cccc} 
\hline\hline
 Line & $\psi$ & $\phi$ & V$_\text{rad}$ (km s$^{-1}$)\\
\hline
He I 7065 \AA&0.31&0.01&$-149\pm$25\\&0.33&0.17&$-196\pm$25\\&0.33&0.24&$-285\pm$25\\&0.34&0.32&$-243\pm$25\\&0.34&0.40&$-265\pm$25\\&0.35&0.47&$-209\pm$25\\&0.36&0.55&$-172\pm$25\\&0.36&0.63&$-189\pm$25\\&0.66&0.67&$-149\pm$7\\&0.37&0.70&$-160\pm$25\\&0.38&0.86&$-152\pm$25\\&0.39&0.93&$-195\pm$25\\
\\
He I 5875 \AA&0.31&0.01&$-136\pm$25\\&0.33&0.17&$-168\pm$25\\&0.33&0.24&$-232\pm$25\\&0.34&0.32&$-247\pm$25\\&0.34&0.40&$-264\pm$25\\&0.35&0.47&$-211\pm$25\\&0.36&0.55&$-203\pm$25\\&0.36&0.63&$-190\pm$25\\&0.66&0.67&$-164\pm$7\\&0.37&0.70&$-164\pm$25\\&0.38&0.86&$-142\pm$25\\&0.39&0.93&$-163\pm$25\\
\\
He I 6678 \AA&0.31&0.01&$-39\pm$25\\&0.33&0.24&$-250\pm$25\\&0.34&0.32&$-235\pm$25\\&0.34&0.40&$-223\pm$25\\&0.35&0.47&$-201\pm$25\\&0.36&0.55&$-161\pm$25\\&0.36&0.63&$-148\pm$25\\&0.38&0.86&$-43\pm$25\\
\hline
\end{tabular}
\end{table}
\begin{table}
\caption{Measured radial velocities of the  O I 7774, 8446 \AA\ transitions. All the velocities have been obtained from the EMMI spectra and XSHOOTER spectra. The error on the velocities is the resolution per pixel of the instrument (25 km s$^{-1}$ or 7 km s$^{-1}$, depending on the spectrograph) apart from the cross correlation points, with an error of 3.5 km s$^{-1}$, due to the algorithm.}
\label{table:vradOI}
\centering
\begin{tabular}{cccc} 
\hline\hline
 Line & $\psi$ & $\phi$ & V$_\text{rad}$ (km s$^{-1}$)\\
\hline
O I 7774 \AA&0.31&0.01&$-39\pm$25\\&0.33&0.09&$-13\pm$25\\&0.33&0.10&25$\pm$25\\&0.34&0.24&$-216\pm$25\\&0.34&0.32&$-209\pm$25\\&0.34&0.40&$-232\pm$25\\&0.35&0.47&$-174\pm$25\\&0.36&0.55&$-170\pm$25\\&0.36&0.63&$-128\pm$25\\&0.66&0.67&$-94\pm$7\\&0.37&0.70&$-42\pm$25\\&0.30&0.78&$-46\pm$25\\&0.38&0.86&$-40\pm$25\\&0.39&0.93&$-24\pm$25\\
\\
O I 8446 \AA\  Cross &0.03&0.08&13.91$\pm$3.5\\&0.04&0.16&11.51$\pm$3.5\\&0.05&0.23&7.1$\pm$3.5\\&0.05&0.31&1.6$\pm$3.5\\&0.06&0.38&$-15.6\pm$3.5\\&0.09&0.77&$-7.0\pm$3.5\\&0.10&0.84&$-5.1\pm$3.5\\&0.10&0.92&4.5$\pm$3.5\\
\\
O I 8446 \AA&0.31&0.01&59$\pm$25\\&0.32&0.10&77$\pm$25\\&0.33&0.17&112$\pm$25\\&0.33&0.24&78$\pm$25\\&0.34&0.36&57$\pm$18\\&0.35&0.47&38$\pm$25\\&0.36&0.55&35$\pm$18\\&0.36&0.67&29$\pm$18\\&0.30&0.78&67$\pm$25\\&0.38&0.88&30$\pm$18\\
\hline
\end{tabular}
\end{table}
\begin{table*}
\caption{Measured polarization values in the B filter. The Stokes parameters' errors are the same of dP$_B$.}
\label{table:buncorr}
\centering
\begin{tabular}{ccccccc} 
\hline\hline
q$_{B}$($\%$) & u$_{B}$( $\%$ ) & P$_{B}$( $\%$ ) & PA$_{B}$ ($^{\circ}$) & JD$-2457800$ & $\psi$ & $\phi$ \\
\hline
1.79 & $-0.35$ & 1.82$\pm$0.20 & 174.5$\pm$3.1 & 516.04& 0.26 & 0.86\\ 2.38 & $-0.15$ & 2.38$\pm$0.21 & 178.2$\pm$2.5 & 517.05 & 0.26 & 0.94\\ 1.73 & $-2.50$ & 3.04$\pm$0.34 & 152.4$\pm$3.2 & 519.10& 0.28 & 0.09\\ 2.75 & $-0.43$ & 2.78$\pm$0.20 & 175.5$\pm$2.0 &522.05& 0.29 & 0.32\\ 2.02 & $-0.51$ & 2.08$\pm$0.10 & 172.9$\pm$1.3 & 523.05& 0.30 & 0.39\\ 2.83 & $-1.27$ & 3.10$\pm$0.38 & 167.9$\pm$3.5 & 526.03 & 0.32 & 0.62\\ 2.41 & $-1.62$ & 2.91$\pm$0.63 & 163.1$\pm$6.1 & 529.03& 0.34 & 0.85\\ 2.68 & $-1.01$ & 2.86$\pm$0.27 & 169.6$\pm$2.7 & 534.02 & 0.37 & 0.23\\ 2.29 & $-0.87$ & 2.45$\pm$0.30 & 169.6$\pm$3.5 &536.01 & 0.38 & 0.39\\ 2.67 & $-1.57$ & 3.10$\pm$0.40 & 164.8$\pm$3.6 & 536.98 & 0.39 & 0.46\\ 2.61 & $-0.64$ & 2.69$\pm$0.27 & 173.1$\pm$2.9 & 538.00 & 0.39 & 0.54\\ 2.65 & $-0.21$ & 2.66$\pm$0.23 & 177.7$\pm$2.5 & 539.01 & 0.40 & 0.61\\ 2.90 & $-1.10$ & 3.10$\pm$0.38 & 169.6$\pm$3.5 & 541.98 & 0.42&0.84\\ 2.05 & $-0.74$ & 2.18$\pm$0.37 & 170.1$\pm$4.8 & 542.99 & 0.42 & 0.92\\ 1.10 & $-1.79$ & 2.10$\pm$0.52 & 150.8$\pm$6.9 & 543.98 & 0.43 & 1.00\\ 1.86 & $-1.80$ & 2.59$\pm$0.36 & 157.9$\pm$3.9 & 544.96 & 0.44 & 0.07\\ 1.90 & $-1.06$ & 2.17$\pm$0.35 & 165.4$\pm$4.6 & 547.98 & 0.45 & 0.30\\ 2.30 & $-0.55$ & 2.36$\pm$0.22 & 173.3$\pm$2.6 & 548.95 & 0.46 & 0.38\\ 2.69 & $-0.31$ & 2.71$\pm$0.25 & 176.7$\pm$2.7 & 551.95 & 0.48 & 0.60\\
\hline
\end{tabular}
\end{table*}
\begin{table*}
\centering
\caption{Same as Table~\ref{table:buncorr} for the V filter.}
\label{table:vuncorr}
\begin{tabular}{cccccccc} 
\hline\hline
q$_\text{V}$($\%$) & u$_\text{V}$( $\%$ ) & P$_\text{V}$( $\%$ ) &PA$_\text{V}$ ($^{\circ}$) &JD$-2457800$ & $\psi$ & $\phi$  \\
\hline
1.81 & $-0.42$ & 1.86$\pm$0.07 & 173.5$\pm$1.1 &516.04 & 0.26 & 0.86\\ 1.93 & $-0.19$ & 1.93$\pm$0.06 & 177.2$\pm$0.9 & 517.05 & 0.26 & 0.94\\ 1.44 & $-1.51$ & 2.08$\pm$0.09 & 156.8$\pm$1.2 &519.10 & 0.28 & 0.09\\ 2.19 & $-0.52$ & 2.25$\pm$0.08 & 173.3$\pm$1.0 &522.05& 0.29 & 0.32\\ 1.65 & $-0.66$ & 1.78$\pm$0.03 & 169.1$\pm$0.5 & 523.05 & 0.30 & 0.39\\ 1.61 & $-1.15$ & 1.98$\pm$0.05 & 162.3$\pm$0.7 & 526.03 & 0.32 & 0.62\\ 1.45 & $-1.16$ & 1.86$\pm$0.10 & 160.7$\pm$1.6 & 529.03 & 0.34 & 0.85\\ 1.95 & $-1.09$ & 2.24$\pm$0.09 & 165.4$\pm$1.1 & 534.02 & 0.37 & 0.23\\ 1.83 & $-0.89$ & 2.03$\pm$0.08 & 167.1$\pm$1.1 & 536.01 & 0.38 & 0.39\\ 1.70 & $-0.85$ & 1.90$\pm$0.10 & 166.7$\pm$1.5 & 536.98 & 0.39 & 0.46\\ 1.58 & $-0.96$ & 1.85$\pm$0.07 & 164.4$\pm$1.2 & 538.00 & 0.39 & 0.54\\ 1.83 & $-0.88$ & 2.03$\pm$0.06 & 167.1$\pm$0.8 & 539.01 & 0.40 & 0.61\\ 1.84 & $-0.64$ & 1.95$\pm$0.10 & 170.4$\pm$1.4 &541.98 & 0.42 & 0.84\\ 1.32 & $-0.84$ & 1.56$\pm$0.09 & 163.7$\pm$1.6 & 542.99 & 0.42 & 0.92\\ 1.33 & $-1.37$ & 1.91$\pm$0.09 & 157.1$\pm$1.3 & 543.98 & 0.43 & 1.00\\ 1.57 & $-1.55$ & 2.20$\pm$0.12 & 157.7$\pm$1.6 & 544.96 & 0.44 & 0.07\\ 1.93 & $-0.82$ & 2.10$\pm$0.08 & 168.5$\pm$1.0 & 547.98 & 0.45 & 0.30\\ 1.81 & $-0.86$ & 2.00$\pm$0.06 & 167.3$\pm$0.9 & 548.95 & 0.46 & 0.38\\ 2.09 & $-0.43$ & 2.13$\pm$0.05 & 174.1$\pm$0.6 & 551.95 & 0.48 & 0.60\\
\hline
\end{tabular}
\end{table*}
\begin{table*}
\caption{Same as Table~\ref{table:buncorr} for the R filter.}
\label{table:runcorr}
\centering
\begin{tabular}{cccccccc} 
\hline\hline
 q$_\text{R}$($\%$) & u$_\text{R}$( $\%$ ) & P$_\text{R}$( $\%$ ) & PA$_\text{R}$ ($^{\circ}$) & JD$-2457800$ & $\psi$ & $\phi$\\
\hline
2.16 & $-0.21$ & 2.17$\pm$0.04 & 177.3$\pm$0.5 & 516.04 & 0.26 & 0.86\\ 2.27 & 0.10 & 2.28$\pm$0.03 & 1.2$\pm$0.3 & 517.05 & 0.26 & 0.94\\ 1.75 & $-1.11$ & 2.07$\pm$0.04 & 163.8$\pm$0.5 & 519.10 & 0.28 & 0.09\\ 2.41 & $-0.31$ & 2.43$\pm$0.02 & 176.4$\pm$0.3 & 522.05 & 0.29 & 0.32\\ 2.07 &$-0.38$ & 2.10$\pm$0.02 & 174.7$\pm$0.3 &523.05 & 0.30 & 0.39\\ 2.06 & $-0.76$ & 2.19$\pm$0.03 & 169.8$\pm$0.4 &526.03 & 0.32 & 0.62\\ 1.91 & $-0.89$ & 2.11$\pm$0.03 & 167.5$\pm$0.5 & 529.03 & 0.34 & 0.85\\ 2.13 & $-0.54$ & 2.19$\pm$0.04 & 172.9$\pm$0.5 & 534.02 & 0.37 & 0.23\\ 2.19 & $-0.48$ & 2.24$\pm$0.03 & 173.9$\pm$0.3 & 536.01 & 0.38 & 0.39\\ 2.10 & $-0.47$ & 2.15$\pm$0.04 & 173.7$\pm$0.6 & 536.98 & 0.39 & 0.46\\ 2.20 & $-0.54$ & 2.26$\pm$0.03 & 173.2$\pm$0.3 & 538.00 & 0.39 & 0.54\\ 2.27 & $-0.56$ & 2.34$\pm$0.03 & 173.0$\pm$0.3 & 539.01 & 0.40 & 0.61\\ 2.11 & $-0.41$ & 2.15$\pm$0.04 & 174.5$\pm$0.5 & 541.98 & 0.42 & 0.84\\ 2.10 & $-0.59$ & 2.18$\pm$0.03 & 172.1$\pm$0.4 & 542.99 & 0.42 & 0.92\\ 1.97 & $-0.76$ & 2.11$\pm$0.03 & 169.5$\pm$0.4 & 543.98 & 0.43 & 1.00\\ 2.05 & $-0.89$ & 2.24$\pm$0.05 & 168.2$\pm$0.6 & 544.96 & 0.44 & 0.07\\ 2.17 & $-0.50$ & 2.23$\pm$0.03 & 173.5$\pm$0.4 & 547.98 & 0.45 & 0.30\\ 2.11 & $-0.56$ & 2.19$\pm$0.03 & 172.6$\pm$0.3 & 548.95 & 0.46 & 0.38\\ 2.41 & $-0.31$ & 2.43$\pm$0.02 & 176.3$\pm$0.2 &551.95& 0.48 & 0.60 \\
\hline
\end{tabular}
\end{table*}
\end{appendix}
\end{document}